\documentclass[11 pt, draftclsnofoot, journal, onecolumn]{IEEEtran}

\usepackage{graphicx,amssymb,amstext,amsmath,color}
\DeclareGraphicsExtensions{.eps,.pdf,.png,.jpg,.gif,.jpeg,.pstex}

\input{epsf.sty}

\usepackage{times, epsf}
\usepackage{amsmath,amssymb}
\usepackage[hyphens]{url}
\usepackage[font = footnotesize]{caption}
\usepackage{ float}
\usepackage{algorithm}
\usepackage{algorithmic}
\usepackage{graphicx}
\usepackage{subfig}
\usepackage{cite}
\usepackage{multirow,tabularx}
\usepackage[pdftex]{epsfig}
\input{epsf.sty}

\usepackage{stackrel}
\usepackage{amsfonts}
\input{epsf.sty}

\usepackage{times, epsf}
\usepackage{graphicx}
\usepackage{cite}
\usepackage{multirow,tabularx}
\usepackage{epstopdf}

\newtheorem{thm}{Theorem}
\newtheorem{lemma}{Lemma}
\newtheorem{cor}{Corollary}
\newtheorem{remark}{\indent \bf Remark}

\newcommand{\expect}[1]{{\mathbb{E}}\left[{#1}\right]}

\newcommand{\E}[1]{\expect{#1}}

\newcommand{\ba}{\begin{array}}
\newcommand{\ea}{\end{array}}

\def\PARstart#1#2{\begingroup\def\par{\endgraf\endgroup\lineskiplimit=0pt}
    \setbox2=\hbox{\uppercase{#2} }\newdimen\tmpht \tmpht \ht2
    \advance\tmpht by \baselineskip\font\hhuge=cmr10 at \tmpht
    \setbox1=\hbox{{\hhuge #1}}
    \count7=\tmpht \count8=\ht1\divide\count8 by 1000 \divide\count7 by\count8
    \tmpht=.001\tmpht\multiply\tmpht by \count7\font\hhuge=cmr10 at \tmpht
    \setbox1=\hbox{{\hhuge #1}} \noindent \hangindent1.05\wd1
    \hangafter=-2 {\hskip-\hangindent \lower1\ht1\hbox{\raise1.0\ht2\copy1}%
    \kern-0\wd1}\copy2\lineskiplimit=-1000pt}

\newcommand{\bit}{\begin{itemize}}
\newcommand{\eit}{\end{itemize}}

\newcommand{\twopartdef}[4]
{
	\left\{
		\begin{array}{ll}
			#1 & \mbox{if } #2 \\
			#3 & \mbox{if } #4 
		\end{array}
	\right.
}

\long\def\comment#1{}

\newfont{\bbb}{msbm10 scaled 700}

\newfont{\bb}{msbm10 scaled 1100}

\newcommand{\ZZ}{\mbox{\bb Z}}

\newcommand{\cv}{{\bf c}}

\newcommand{\xv}{{\bf x}}

\newcommand{\Bm}{{\bf B}}

\newcommand{\Dm}{{\bf D}}

\newcommand{\Gm}{{\bf G}}

\newcommand{\Qm}{{\bf Q}}

\newcommand{\Vm}{{\bf V}}

\newcommand{\Ec}{{\cal E}}

\newcommand{\Gc}{{\cal G}}
\newcommand{\Hc}{{\cal H}}

\newcommand{\Kc}{{\cal K}}

\newcommand{\Nc}{{\cal N}}

\newcommand{\Rc}{{\cal R}}
\newcommand{\Sc}{{\cal S}}

\newcommand{\Uc}{{\cal U}}

\newcommand{\gammav}{\hbox{\boldmath$\gamma$}}

\newcommand{\muv}{\hbox{\boldmath$\mu$}}

\newcommand{\Thetav}{\hbox{\boldmath$\Theta$}}

\newcommand{\xiv}{\hbox{\boldmath$\xi$}}

\newcommand{\kappav}{\hbox{\boldmath$\kappa$}}

\newcommand{\Thetam}{\hbox{\boldmath$\Theta$}}

\newcommand{\Xim}{\hbox{\boldmath$\Xi$}}

\newcommand{\herm}{{\sf H}}

\newcommand{\transp}{{\sf T}}

\begin{document}
\title{WiFlix: Adaptive Video Streaming in Massive MU-MIMO Wireless Networks}
\author{\IEEEauthorblockN{D. Bethanabhotla, G. Caire and M. J. Neely}
\thanks{All authors are with the Ming Hsieh Department of Electrical Engineering, Viterbi School of Engineering, 
University of Southern California, Los Angeles CA 90089. Email: {\tt bethanab, caire, mjneely@usc.edu}}\\
\thanks{This work was partially supported by NSF Grant CCF-1423140 and NSF Grant CCF-0747525.}
}

\maketitle

\vspace{-2 cm}


\begin{abstract} 
We consider the problem of simultaneous on-demand streaming of stored video to multiple users in a multi-cell wireless network where multiple unicast streaming sessions are run in parallel and share the same frequency band. Each streaming session is formed by the sequential transmission of video ``chunks'', such that each chunk arrives into the corresponding user playback buffer within its playback deadline. We formulate the problem as a Network Utility Maximization (NUM) where the objective is to fairly maximize users' video streaming Quality of Experience (QoE) 
and then derive an iterative control policy using Lyapunov Optimization, which solves the NUM problem up to any level of accuracy and yields an {\em online protocol} with control actions at every iteration decomposing into two {\it layers} interconnected by the users' {\it request queues} : i) a video streaming adaptation layer reminiscent of DASH, implemented at each user node; ii) a transmission scheduling layer where a max-weight scheduler is implemented at each base station. The proposed chunk request scheme is a {\it pull} strategy where every user opportunistically requests video chunks from the neighboring base stations and dynamically adapts the quality of its requests based on the current size of the request queue. For the transmission scheduling component, we first describe the general max-weight scheduler and then particularize it to a wireless network where the base stations have multiuser MIMO (MU-MIMO) beamforming capabilities. We exploit the {\it channel hardening} effect of large-dimensional MIMO channels (massive MIMO) and devise a low complexity user selection scheme to solve the underlying combinatorial problem of selecting user subsets for downlink beamforming, which can be easily implemented and run independently at each base station. Further, through simulations, we show that deploying MU-MIMO significantly improves video streaming performance and also that the proposed cross-layer approach is able to serve users more fairly than a baseline scheme representative of current systems running independently designed protocol layers. 
\end{abstract}


\section{Introduction}  \label{sec-intro-wiflix}

Demand for video content over wireless networks has grown dramatically in recent years and it is predicted to account for 69\% of the total mobile data traffic by 2018 \cite{Cisco1}. 
This  is mainly due to on-demand video streaming, enabled by multimedia devices such as tablets and smartphones. 
In addition, recent measurement studies~\cite{conviva} reveal that, in 2013, around 26.9\% of video streaming sessions on the Internet experienced 
playback interruption due to re-buffering, 43.3\% were impacted by low resolution, and 4.8\% failed to start altogether. 
At the application layer, {\em Dynamic Adaptive Streaming over HTTP} (DASH) \cite{Dash,begen2011watching}\footnote{This includes industry products such as 
Microsoft Smooth Streaming and Apple HTTP Live Streaming, which qualitatively work in the way assumed in our paper, up to minor variations which are irrelevant for the 
present theoretical treatment.}  has become a de-facto industry standard approach to handle 
video streaming over wireless networks. In DASH, each user (client) monitors the available capacity during a video streaming session 
and chooses adaptively and dynamically the most appropriate video quality level correspondingly. The video files are divided into ``chunks'', which are downloaded by sequential HTTP requests. 
Different quality levels can be obtained either by storing multiple versions of the same video encoded at different bit-rates, 
or by using scalable video coding and sending an adaptive number of refinement layers\cite{sanchez2011idash}.

%

In this way, DASH attempts to maintain a reasonable quality of experience (QoE) even under changing network conditions. 
However, operating at the application layer only is not sufficient to achieve a fully satisfactory performance. 
For instance, popular video platforms such as Youtube and Netflix, which employ DASH at the application layer, have realized this fact and recently released 
Video Quality Reports~\cite{googlevidqoe, netflixvidqoe} where they compare and contrast different network service providers (ISP) in a given geographical area and rank/label them 
as either Lower Definition (LD) or Standard Definition (SD) or High Definition (HD) based on the quality of video streaming activity in their network over a certain time frame
in order to inform users that the choice of ISP can affect video streaming QoE. 

\subsection{Motivation and related work}\label{subsec-motivation-wiflix}

In order to cope with this problem, a {\it cross layer} optimization approach has been proposed in several works (e.g., see \cite{joseph2013nova, wang2014adaptive, khalek2013loss, zhao2014scheduling,chen2011adaptive, bethanabhotla2013joint,miller2015control}). In these works, the video streaming QoE is defined in terms of performance metrics such as 
video quality, probability of stall events (i.e., when the playback buffer is empty and video playback stops), 
pre-buffering time, and re-buffering time. However, the joint optimization of these metrics by directly controlling the dynamics of the playback buffers 
of all the users in the network requires solving a Markov Decision Problem (MDP) which is typically quite difficult and incurs the well-known curse of dimensionality. 
For instance, \cite{wang2014adaptive} considers adaptive video transmission in a much simpler setting of a point-to-point wireless link and formulates the problem as an MDP which is then solved using 
the value iteration algorithm. However, even in such a simple point-to-point scenario, the value iteration policy requires extensive computation to be done offline and stored in a lookup table 
which is then used for the actual transmission. On the other hand, the work \cite{bethanabhotla2013joint} takes a cross-layer approach and considers video delivery in the general case of a 
multiuser wireless network where users are served by wireless {\em helper nodes}.\footnote{Our treatment applies, at a very high level, to any infrastructure-based wireless network such 
as conventional cellular, small cells, WLAN, and heterogeneous compositions thereof (e.g., a cellular network with wifi off-load). Therefore, throughout this paper, we refer to infrastructure nodes simply as ``helpers''.} 
In order to obtain a tractable formulation for the multiuser network,  \cite{bethanabhotla2013joint} adopts a ``divide and conquer" approach where first 
the problem of maximizing a function of the time-averaged video qualities, subject to queue stability is solved, and then the delay jitter is taken care of by appropriately dimensioning the pre-buffering 
and re-buffering times, exploiting the fact that the playback buffer can absorb the delay fluctuations around the (bounded) mean.  However, in \cite{bethanabhotla2013joint} a ``push" scheduling policy 
is considered, for which video chunks can be served out of order and may result in data loss in the presence of intermittent connectivity and/or mobility. 
In this paper, we fix this problem and introduce a new ``pull'' strategy, that is robust to fast topology variations. 
Our scheme allows each user to opportunistically pull data always in the correct sequential order from neighboring helper nodes. 
This results in smoother and more reliable performance. Another shortcoming of \cite{bethanabhotla2013joint} is that it considers only helpers operating according to OFDM/TDMA, i.e., serving 
at most one user per transmission resource (referred to as {\it PHY frame} hereafter). 
As a matter of fact, the current wireless technology trend is rapidly evolving towards multiuser MIMO (MU-MIMO) schemes ( e.g., see \cite{802.11_web, andrews5g, bethanabhotla2014optimal, pcellwhite}) 
where multiple users can be served on the same PHY frame by spatial multiplexing. The current work therefore allows for general wireless channel models, including MU-MIMO as a special case.

\subsection{Contributions}

Motivated by the above considerations, this paper focuses on the problem of dynamic adaptive video streaming in a wireless network formed by a number of densely deployed 
wireless helper nodes serving multiple wireless users over a given geographic coverage area and on the same shared channel bandwidth. 
We address the problem by {\em jointly optimizing} the video quality adaptation at the DASH layer (application layer) and the transmission scheduling of users at the PHY/MAC layer. 
This is obtained through a  cross layer approach where the appropriate queue sizes maintained at the users act as a bridge between the layers. 
In particular, the novel contributions of this paper are as follows:
\begin{itemize}
\item We introduce the notion of a {\it request queue}. This is a virtual queue, maintained by each user, that serves to sequentially request video chunks from helper nodes, such that the choice of the helper node and the quality at which each video chunk is requested can be adaptively adjusted. Each user, upon deciding the quality of the chunk, {\it requests} the bits corresponding to that chunk and places them in the request queue. Note that this does not mean the user has already downloaded the chunk, but the chunk bits are ``virtually" placed in the request queue and will be taken out when the chunk is effectively delivered to the user. In this way, the user maintains in the request queue all the chunk bits  that have been requested but not downloaded and adaptively adjusts the quality of future chunk requests based on its size. In addition, the user broadcasts this size to the helpers in its current vicinity and  ``pulls" bits from them in the right order necessary for video playback starting at the {\em Head Of Line} (HOL) of the request queue. Even if a mobile user gets out of range of a helper while downloading the HOL bits, it can still re-request those bits from the new helper in its current vicinity. In this way, the user always downloads chunks in the playback order and does not skip any of them. 
This improves significantly upon the ``push" scheme proposed in~\cite{bethanabhotla2013joint} where the chunks could be downloaded out of order due to different transmission queue delays at different helper nodes, 
or skipped if a user moves out of a helper's coverage after placing a request.

\item We systematically obtain our cross-layer policy as the dynamic solution of a Network Utility Maximization (NUM) problem, where the network utility function is given in terms of the 
users' time-averaged video quality, and the maximization constraints are given by imposing stability of each request queue. The stability constraint implies that every requested chunk will be eventually delivered, 
while delivery in the right sequential order is guaranteed by the request queue mechanism described above. 
The proposed policy decomposes naturally into two interconnected {\it layers}: i) a video streaming adaptation layer reminiscent of DASH, implemented at each user node, and involving the 
adaptive video quality selection and placement of the video chunk requests into the request queue; ii) a transmission scheduling layer where a max-weight scheduler 
is implemented at the helpers. These two layers are interconnected by the users' {\it request queues}, which form the weights for the max-weight scheduler. 
Although queue stability guarantees that all requested chunks are eventually delivered, such delivery may still occur, occasionally, after the corresponding playback deadline. 
In this case, we are in the presence of a stall event. In order to control the stall event probability and make it sufficiently small, we follow the same divide and conquer approach
of \cite{bethanabhotla2013joint}, and adaptively set the pre-buffering/re-buffering time by monitoring the chunk delivery delay in a sliding window. This approach has the advantage of 
yielding very good performances also in terms of stall event probability, while allowing for the elegant and mathematically tractable NUM framework in terms of the video quality maximization. 

\item  We particularize the max-weight transmission policy to a network of helpers with MU-MIMO capabilities, 
where the scheduling actions consist of choosing the subset of users for MU-MIMO beamforming at each helper. 
By exploiting the ``channel hardening" effect of large-dimensional MIMO channels (massive MIMO)  \cite{Marzetta-TWC10, Huh11,hoydis2011massive}, 
we reduce the combinatorial weighted sum rate maximization over the multiuser multicell network (which would involve an exponentially complex exhaustive user selection, or some polynomial complexity 
heuristic greedy user selection at each helper) to a simple subset selection problem which is optimally solved by a low complexity algorithm. 
The algorithm can be implemented independently at the MAC layer of each helper. The only information that needs to be exchanged between the layers is the length of the users' request queues, which
can be easily gathered as ``protocol information'' via the uplink, together with the chunk requests. 

\item We show through simulation in a realistic network topology and using actual encoded video data that the proposed system is very effective in improving the average 
video quality and reducing the percentage of time spent in buffering mode.
 
\end{itemize}

\section{System Model} \label{sec-sysmodel-wiflix}

We consider a wireless network with multiple users and multiple helper stations sharing the same bandwidth. The network is defined by a bipartite graph $\Gc = (\Uc, \Hc, \Ec)$, where $\Uc$ denotes the set of users, $\Hc$ denotes the set of helpers, and $\Ec$ contains edges for all pairs $(h,u)$ such that helper $h$ can transmit information to user $u$. We denote by $\Nc(u) \subseteq \Hc$ the neighborhood of user $u$, i.e., $\Nc(u) = \{ h \in \Hc : (h,u) \in \Ec\}$. Similarly,  $\Nc(h) = \{u \in \Uc : (h,u) \in \Ec\}$.  Each user $u \in \Uc$ requests a video file $f_u$ which is formed by a sequence of  chunks. Each chunk corresponds to a group of pictures (GOP) that are encoded and decoded as stand-alone units~\cite{sanchez2011idash}. Chunks have a fixed playback duration, given by $T_{\rm gop} = \mbox{(\# frames per GOP)}/\eta$, where $\eta$ is the frame rate, expressed in frames per second. The streaming process consists of transferring chunks from the helpers to the requesting users such that the playback buffer at each user contains the required chunks at the beginning of each chunk playback deadline. The playback starts after a short pre-buffering time, during which the playback buffer is filled by a determined amount of ordered chunks. The details related to pre-buffering and chunk playback deadlines are discussed in Section \ref{sec-prebuffering-wiflix}.

Each file $f$ is encoded at a finite number of different quality/compression levels $m \in \{1, \ldots, N_f\}$ \cite{begen2011watching}. 
Due to the variable bit rate (VBR) nature of video coding\cite{ortega2000variable}, the quality-rate profile of a given file $f$ may vary from chunk to chunk. 
In particular, we let $D_f(m,i)$ and $B_f(m,i)$ denote the video quality measure (e.g., see \cite{wang2004image}) and the size (in number of bits) 
of the $i$-th chunk in file $f$ at quality level $m$, respectively.  

\subsection{Time-scales}

It is important to note that the time scale at which chunks are requested and the time scale at which PHY layer transmissions are scheduled differ by $1-3$ orders of magnitude. For instance, in current video streaming technology \cite{Dash}, the typical video chunk spans a duration of $0.5-2$ seconds while the duration of a {\it PHY frame} is of the order of milliseconds.\footnote{For example, with a PHY frame duration of $10$ ms (as in the LTE 4G standard \cite{molisch2010wireless}) and assuming $T_{\rm gop}=0.5$s, a video chunk spans $n = \frac{0.5}{10 \cdot 10^{-3}} = 50$ PHY frames.} 
In the following, we consider dynamic scheduling policies that operate at the PHY frame time scale, i.e., they provide a scheduling/resource allocation decision at each PHY frame times $t \in \ZZ$. 
However, new chunks are requested at multiples of the chunk time, i.e., at times $t = in$ for $i \in \ZZ$ and $n$ denoting the number of PHY frames per chunk time, assumed here to be an integer for simplicity. 
In the rest of the paper, we will use consistently the following notation:  index $t$ denotes the PHY frame {\it transmission slots},  and the index $i$ denotes {\it video chunks}. 


\subsection{Request Queue Dynamics}

At the beginning of the $i$-th chunk time, each user $u \in\Uc$ requests a particular quality mode for the $i$-th chunk of its video stream. 
That is, on each slot $t \in \{0, n, 2n, 3n, \ldots\}$, each user $u \in \Uc$ specifies the quality mode $m_u(t) \in \{ 1, 2, \ldots, N_{f_u}\}$ for its next video chunk. 
This decision specifies the quality $D_{f_u}(m_u(t), t)$ and the amount of bits $B_{f_u}(m_u(t), t)$ associated with that chunk.  
As these decisions are made only at times $t$ that are multiples of $n$, it is convenient to define: 
\begin{equation}\label{zero-def-wiflix}
D_{f_u}(m_u(t), t) = 0 ~\mathrm{and}~ B_{f_u}(m_u(t),t) = 0 ~\mathrm{for}~ t \not \in \{0, n, 2n, \ldots \}. 
\end{equation}
The bits $B_{f_u}(m_u(t), t)$ are called the {\em requested bits} of user $u$ at slot $t$, and are placed in a \emph{request queue} $Q_u(t)$. The request queue evolves over the transmission slots $t \in \{0, 1, 2,\ldots \}$ as:  
\begin{align}
\label{q-update-wiflix}
Q_{u}(t+1)=\max\{Q_{u}(t)-\mu_{u}(t)+B_{f_u}(m_u(t),t), 0\} \;\;\;\; \forall~ u \in \Uc,
\end{align}
where $\mu_u(t)$ is the amount of bits downloaded by user $u$ on slot $t$. Note that the request queue in (\ref{q-update-wiflix}) can decrease every transmission slot $t$ as new bits are downloaded, 
but can only increase on slots $t = in$, i.e., on integer multiples of $n$.  
Intuitively, $Q_u(t)$ consists of bits associated with all chunks that have been requested by user $u$ but not yet 
fully received. 

The quantity  $\mu_{u}(t)$ indicates the instantaneous aggregate downloading rate of user $u$ on slot $t$, expressed in bits per slot. This is given by 
\begin{equation}\label{total-bits-wiflix}
\mu_{u}(t) = \sum_{h \in \Nc(u)}\mu_{hu}(t)1_{hu}(t)
\end{equation} 
where $1_{hu}$ is an indicator function, equal to 1 if helper $h$ has the video file requested by user $u$ and zero otherwise,  
and $\mu_{hu}(t)$ is the rate served by helper $h$ to user $u$ on slot $t$. The matrix $[\mu_{hu}(t)]$ of transmission rates is selected within a set of feasible transmission rate matrices for slot $t$. 
The set of all rate matrices supported by the network at a given slot time $t$ is referred to as the {\em feasible instantaneous rate region} at time $t$, and depends on 
the network topology and channel state (e.g., on the fading channels realization). 
Specifically, let $\omega(t)$ represent the topology state on slot $t$, being a vector of parameters that affect transmission, such as current device locations and/or channel conditions. 
Assume $\omega(t)$ takes values in an abstract set $\Omega$, possibly being an infinite set. 
For each $\omega \in \Omega$, define $\Rc(\omega)$ to be the feasible rate region of the network for state $\omega$. Then, the feasible instantaneous rate region is $\Rc(\omega(t))$. 
For example, the set $\Rc(\omega)$ may include the constraint that each user can receive a positive rate from at most one helper and/or constrain helpers to restrict transmissions to at most 
$S$ users, where $S$ denotes the maximum number of downlink data streams (spatial multiplexing gain) that the helper station can handle.\footnote{See~\cite{neely2012wireless} for a discussion of various wireless multiple access scenarios and interference models that fit this general framework.} 
The set $\Rc(\omega)$ can also handle models that allow simultaneous download from multiple helpers (for instance, in a cellular CDMA system with macro diversity), or information-theoretic capacity regions of
various network topology models, inclusive of broadcast and interference constraints (e.g., \cite{el2010lecture}). We also mention here that this framework can also handle non-wireless scenarios. 
For example, it can constrain $[\mu_{hu}(t)]$ to be permutation matrices associated with packet switch constraints. 
However, as explained in Section \ref{sec-intro-wiflix}, it is desirable for current and future systems to take advantage of massive MU-MIMO capabilities at the helpers. 
Section \ref{sec-channel-model-wiflix} specifies $\Rc(\omega)$ for the relevant wireless scenario with helpers employing massive MU-MIMO, which is the primary focus of this paper. 
The simulation results in Section \ref{sec-simul-wiflix} are carried out under this specific wireless model.
 
 \begin{remark}
 {\em Each user $u$ maintains $Q_u(t)$ and updates it according to (\ref{q-update-wiflix}) every transmission slot $t$.  A small amount of bookkeeping is also required by the user to associate the bits $Q_u(t)$ with their appropriate chunks.  Specifically, each user maintains a list of chunks it has requested but not yet fully received, along with the quality modes it requested for each chunk.  It can receive new bits on slot $t$ only from a helper that has its requested file, and only if $Q_u(t)>0$.  When downloading these bits, the user first informs the helper of the requested chunks, the desired quality levels, and the bit location needed for downloading the residual bits of the next-in-line chunk. }
\end{remark}

\section{Problem Formulation and Streaming Policy} \label{sec-formulation-wiflix}

When optimizing the users' video QoE we have to take into account that users compete for the same shared transmission resource 
(the network wireless spectrum and the helpers spatial downlink data streams) and, given the fact that the users are placed in arbitrary positions with respect to the helpers, 
their attainable service rates may be quite different. Hence, some fairness criterion must be enforced. 
In addition, we need to carefully define the notion of QoE, since the adaptive nature of the streaming process involves a possibly time-varying quality level across the streaming sessions. 

As already mentioned briefly before, we remark once again that, in order to obtain a tractable formulation, we adopt the {\it divide and conquer approach} pioneered in \cite{bethanabhotla2013joint}:
\begin{enumerate}
\item We first formulate the NUM problem (\ref{NUMproblem-wiflix}), where the network utility function is a concave and component wise non-decreasing function of the time averaged 
users' {\it requested} video quality and the maximization is subject to the stability of all the request queues in the system. 
\item We then solve the NUM problem using the Lyapunov Optimization framework and obtain the {\it drift-plus-penalty} policy which adapts to arbitrarily changing network conditions and in fact 
is optimal (with respect to the NUM problem) under non-stationary and non-ergodic evolution of the underlying network state process.  
\item Since all the request queues in the system are ensured to be stable, the {\it requested} video chunks are eventually {\it delivered}. 
However, in order to ensure that all the video chunks are delivered within their playback deadline, it suffices for every user to choose a {\it pre-buffering} time which exceeds the largest delay with which 
a chunk is delivered. In particular, when the maximum delay of each request queue in the system admits a deterministic upper bound, setting the pre-buffering time larger than such a bound makes the playback buffer under rate zero. However, for a system with arbitrary (non-stationary, non-ergodic) evolution of the underlying network state process (for e.g., arbitrary user mobility and arbitrary per-chunk fluctuations of video coding rate due to VBR coding), such deterministic upper bounds on the maximum delay may not exist or are too loose to be useful in practice. Hence, in Section \ref{sec-prebuffering-wiflix}, we propose a method to locally estimate the delays with which video chunks are delivered, such that each user can calculate its pre-buffering and re-buffering times to be larger than the locally estimated maximum delay. Through simulations in Section \ref{sec-simul-wiflix}, we demonstrate the effectiveness of the combination of the drift-plus-penalty policy and the adaptive pre-buffering scheme.
\end{enumerate}
In the rest of this section, we focus on the NUM problem formulation and its solution through the drift-plus-penalty approach. 
Throughout this work, we use the following notation for the time average quantity of interest:  we let
$\overline{D}_u:=\lim_{t\rightarrow \infty}\frac{1}{t}\sum_{\tau=0}^{t-1}\E{ D_{f_u}\left(m_u(\tau),\tau\right)}$ denote the time average of the expected quality of user $u$, 
and  $\overline{Q}_{u} := \lim_{t\rightarrow \infty}\frac{1}{t}\sum_{\tau=0}^{t-1} \E{Q_{u}\left(\tau\right)}$ to be the time average of the expected length of the request queue at user $u$, 
assuming that these limits exist. More in general, we use the overline notation to indicate limiting time-averages.\footnote{The existence of these limits is assumed temporarily for ease of exposition of the optimization problem (\ref{NUMproblem-wiflix}) but is not required for the derivation of the scheduling policy and for the proof of Theorem \ref{main-result-wiflix}.}
Let $\phi_u(\cdot)$ be a concave, continuous, and non-decreasing function 
defining network utility vs. video quality for user $u \in \Uc$. The NUM problem that we wish to solve is given by:
\begin{subequations}\label{NUMproblem-wiflix}
\begin{align}
\textrm{maximize}  & \;\;\; \sum_{u \in \Uc}\phi_u(\overline{D}_u)  \\
 \textrm{subject to} & \;\;\; \overline{Q}_{u} < \infty~\forall~ u \in \Uc \\
& \;\;\; [\mu_{hu}(t)] \in \Rc(\omega(t)) ~\forall~t \\
& \;\;\; m_u(t) \in \{1, 2, \ldots, N_{f_u} \}~\forall~u \in \Uc,~\forall~t,
\end{align}
\end{subequations}
where requirement of finite $\overline{Q}_{u}$ corresponds to the {\it strong stability} condition for all 
the queues \cite{neely2010stochastic}.

By appropriately choosing the functions $\phi_u(\cdot)$, we can impose some desired notion of fairness. For example, a general class of concave functions suitable for this purpose is given by the $\alpha$-fairness network utility, 
defined by~\cite{mo2000fair} 
\begin{equation}
\phi_u(x) = \left \{ \begin{array}{ll}
\log x & \alpha = 1 \\
\frac{x^{1- \alpha}}{1 - \alpha} & \alpha > 0, \;\; \alpha \neq 1 \end{array} \right .
\end{equation} 
In this case, it is well-known that $\alpha = 0$ yields the maximization of the sum quality (no fairness), 
$\alpha \rightarrow \infty$ yields the maximization of the worst-case quality (max-min fairness) and
$\alpha = 1$ yields the maximization of the geometric mean quality (proportional fairness). 

In order to solve problem (\ref{NUMproblem-wiflix}) using the stochastic optimization theory developed in \cite{neely2010stochastic}, it is convenient to transform it into an equivalent problem that involves  the maximization of a single time average. This transformation is achieved through the use of auxiliary variables $\gamma_u(t)$ and the corresponding virtual queues $\Theta_u(t)$ with buffer evolution:
\begin{align}
\Theta_u(t+1) = \max{\{\Theta_u(t) + \gamma_u(t) - D_{f_u}(m_u(t),t),0 \}}.  \label{virt-update-wiflix}
\end{align} 
Consider the transformed problem:
\begin{subequations}\label{transform-wiflix}
\begin{align}
 \textrm{maximize} & \;\;\; \sum_{u \in \Uc}\overline{\phi_u({\gamma}_u)}\\
 \textrm{subject to} & \;\;\;  \overline{Q}_{u}
< \infty~\forall~ u \in \Uc \\
& \;\;\; \overline{\gamma}_u \leq \overline{D}_u~\forall~u~\in~\Uc  \label{gammaconst-wiflix}\\
& \;\;\; D_u^{\min} \leq \gamma_u(t) \leq D_u^{\max}~\forall~u~\in~\Uc \\
& \;\;\; [\mu_{hu}(t)] \in \Rc(\omega(t)) ~\forall~t \\
& \;\;\; m_u(t) \in \{1, 2, \ldots, N_{f_u} \}~\forall~u \in \Uc,~\forall~t,
\end{align}
\end{subequations}
where $D_u^{\min}$ and $D_u^{\max}$ are uniform lower and upper bounds on the quality 
function $D_{f_u}(\cdot, t)$. Notice that constraints~(\ref{gammaconst-wiflix}) correspond to stability of the virtual queues $\Theta_u$, since $\overline{\gamma}_u$ and $\overline{D}_u$ are the time-averaged arrival rate and the time-averaged service rate for the virtual queue given in (\ref{virt-update-wiflix}). We have: 

\begin{lemma} \label{equivalence-wiflix}
Problems (\ref{NUMproblem-wiflix}) and (\ref{transform-wiflix}) are equivalent. 
\hfill 
\end{lemma}

\begin{IEEEproof} \label{proof-lem-wiflix}
The proof is well-known (see \cite{bethanabhotla2013joint, neely2010stochastic} for instance) and is omitted due to space constraints. 
 \end{IEEEproof}
 
 \subsection{The Drift-Plus-Penalty Expression}  \label{dpp-expression-wiflix}
Let $\Qm(t)$ denote the column vector containing the backlogs of queues $Q_{u}~\forall~u\in\Uc$,  
let $\Thetam(t)$ denote the column vector for the virtual queues $\Theta_u~\forall~u\in\Uc$,
$\gammav(t)$ denote the column vector with elements $\gamma_u(t)~\forall~u\in\Uc$, $\Bm(t)$ denote the column vector with elements $B_{f_u}(m_u(t),t)~\forall~u\in\Uc$, 
$\Dm(t)$ denote the column vector with elements $D_{f_u}(m_u(t),t)~\forall~u\in\Uc$ and $\muv(t)$ denote the column vector with elements $\mu_u(t)~\forall~u \in \Uc$ as defined in (\ref{total-bits-wiflix}). Let $\Gm(t) = \left[ \Qm^\transp(t), \Thetam^\transp(t) \right]^\transp$ be the composite vector of queue backlogs and define the quadratic Lyapunov function $L(\Gm(t)) = \frac{1}{2} \Gm^\transp (t) \Gm(t)$.  Intuitively, taking actions to push $L(\Gm(t))$ down tends to maintain
stability of all queues. Define $\Delta(\Gm(t))$ as the one-slot drift of the Lyapunov function at slot $t$ :
\begin{align}
\Delta(\Gm(t)) \triangleq L(\Gm(t+1))-L(\Gm(t))
\end{align}
The drift-plus-penalty algorithm is designed to observe the queues, the current $B_{f_u}(\cdot, t)$, $D_{f_u}(\cdot,t)$ for all users $u$ and $\omega(t)$ on each slot $t$, and to then choose quality mode $m_u(t)$ for all users $u$, matrix of transmitted bits $(\mu_{hu}(t)) \in\Rc(\omega(t))$ and $\gamma_u(t)$ subject to $D_u^{\min} \leq \gamma_u(t) \leq D_u^{\max}$ to minimize a bound on the following {\it drift-plus-penalty expression}:
\begin{align}
\Delta({\bf G}(t))&-V\sum_{u \in \Uc}\phi_u(\gamma_u(t))
\end{align}
where $V$ is a non-negative weight that affects a performance bound. Intuitively, the value of $V$ affects the extent to which the control actions on slot $t$ emphasize utility maximization in
comparison to drift minimization.
\begin{lemma}
Under any control algorithm, the drift-plus-penalty expression satisfies:
\begin{align}
\label{dpp-ineq-wiflix}
\Delta({\bf G}(t))-V\sum_{u \in \Uc}\phi_u(\gamma_u(t))  \leq \Kc &-V\sum_{u \in \Uc}\phi_u(\gamma_u(t))+\left(\Bm(t)-\muv(t)\right)^\transp{\bf Q}(t) \notag \\
&+\left({\gammav}(t)-\Dm(t)\right)^\transp{\Thetav}(t).
\end{align}
where $\Kc$ is a uniform upper bound on the term $$\frac{1}{2}\left[\left(\Bm(t)-\muv(t)\right)^\transp\left(\Bm(t)-\muv(t)\right)+\\ \left(\gammav(t)-\Dm(t)\right)^\transp\left(\gammav(t)-\Dm(t)\right)\right].$$
\end{lemma}

\begin{IEEEproof}
Expanding the quadratic Lyapunov function, we have
\begin{align}
\label{lyapbound-wiflix}
 &L(\Gm(t+1))-L(\Gm(t)) \notag \\
 & = \frac{1}{2}\left (\Qm^\transp(t+1) \Qm(t+1)- \Qm^\transp (t) \Qm(t) \right )
+ \frac{1}{2} \left ( \Thetam^\transp(t+1) \Thetam(t+1)- \Thetam^\transp (t) \Thetam(t) \right )\notag\\
& =  \frac{1}{2}\left[ \left( \max\{ \Qm(t) - \muv(t) + \Bm(t), {\bf 0} \} \right )^\transp \left(\max\{\Qm(t)- \muv(t) + \Bm(t), {\bf 0} \}\right)
- \Qm^\transp(t) \Qm(t) \right] \notag \\
&~~~~ + \frac{1}{2}\left[\left(\max\{ \Thetam(t)+\gammav(t)-\Dm(t), {\bf 0}\} \right)^\transp\left(\max\{\Thetam(t)+
\gammav(t) - \Dm(t), {\bf 0} \} \right) - \Thetam^\transp(t)\Thetam(t)\right],
\end{align}
where we have used the queue evolution equations (\ref{q-update-wiflix}) and (\ref{virt-update-wiflix}) and ``max'' is applied componentwise.  

Using the fact that for any non-negative scalar quantities $\Theta,\gamma$ and $D$ we have the inequalities
\begin{align}
(\max\{\Theta + \gamma - D, 0\})^2 \leq (\Theta + \gamma -D)^2 = \Theta^2 + (\gamma-D)^2 + 2\Theta(\gamma-D), \label{ineq2-wiflix}
\end{align}
we have
\begin{align}
L({\bf G}(t+1))-L({\bf G}(t)) &\leq  \frac{1}{2}\left(\Bm(t)-\muv(t)\right)^\transp\left(\Bm(t)-\muv(t)\right)+\left(\Bm(t)-{\boldsymbol \mu}(t)\right)^
\transp{\bf Q}(t) \notag \\
&~~~~ + \frac{1}{2}\left(\gammav(t)-\Dm(t)\right)^\transp\left(\gammav(t)-\Dm(t)\right) + \left(\gammav(t)-
\Dm(t)\right)^\transp\Thetav(t) \label{drift-bound-wiflix}
\end{align}

Under the realistic assumption that the chunk sizes, the transmission rates and the video quality measures are bounded above by some constants, independent of $t$, the term 
$$\frac{1}{2}\left[\left(\Bm(t)-\muv(t)\right)^\transp\left(\Bm(t)-\muv(t)\right)+\\ \left(\gammav(t)-\Dm(t)\right)^\transp\left(\gammav(t)-\Dm(t)\right)\right]$$ is bounded above by a constant $\Kc$. Using this fact and adding the penalty term $-V\sum_{u \in \Uc}\phi_u(\gamma_u(t))$ on both sides of the inequality (\ref{drift-bound-wiflix}) yields the result.
\end{IEEEproof}

 The drift-plus-penalty (DPP) policy described below acquires information about the queue states ${\bf G}(t)$, the rate-quality profile $(B_{f_u}(\cdot,t), D_{f_u}(\cdot, t))$ for all users $u$ and  
 the channel state $\omega(t)$ at every slot $t$, and chooses control actions $m_u(t)$, $[\mu_{hu}(t)] \in\Rc(\omega(t))$ and $\gamma_u(t)$,  
 subject to $D_u^{\min} \leq \gamma_u(t) \leq D_u^{\max}$, in order to minimize the last three terms on the right hand side of the inequality (\ref{dpp-ineq-wiflix}). 
 
The non-constant part in the right hand side of (\ref{dpp-ineq-wiflix}) can be re-written as:
\begin{equation}
\left [ \Bm^\transp (t) {\bf Q}(t) - \Dm^\transp(t)\Thetav(t)\right ]  
 - \left [  V\sum_{u \in \Uc}\phi_u(\gamma_u(t)) -  \gammav^\transp(t)\Thetav(t)  \right ]
- {\boldsymbol \mu}^\transp (t){\bf Q}(t).
\label{DPP-wiflix}
\end{equation}
The resulting control actions are given by the minimization, at transmission slot $t$, of the expression in (\ref{DPP-wiflix}). Notice that the first term of (\ref{DPP-wiflix}) depends only on $m_u(t)~\forall~u \in \Uc$, the second term of (\ref{DPP-wiflix}) depends only on $\gammav(t)$ and 
the third term of (\ref{DPP-wiflix}) depends only on $\muv(t)$.
Thus, the overall minimization decomposes into three separate sub-problems, yielding the layered scheme given below.


\subsection{The Drift-Plus-Penalty Policy}\label{dpp-policy-wiflix}


We address the minimization of (\ref{DPP-wiflix}) focusing separately on its (separable) components. 

\subsubsection{Control actions at the user nodes (pull congestion control)}
The first term in (\ref{DPP-wiflix}) is given by 
\begin{align}
\sum_{u \in \Uc} \left \{Q_{u}(t)B_{f_u}(m_u(t),t) - \Theta_u(t) D_{f_u}\left(m_u(t),t\right)\right \}.
\end{align}
The minimization variables $m_u(t)$ appear in separate terms of the sum and hence can be optimized separately over each user $u \in \Uc$. Thus, each user observes the queues $Q_u(t), \Theta_u(t)$ and is aware of the 
the rate-quality profile $(B_{f_u}(\cdot, t), D_{f_u}(\cdot, t))$ on slot $t$ (vidoe meta-data), 
so that it can choose the quality level of the requested chunk at every video chunk slot $i$, i.e., at transmission slots $t \in \{ in: i \in \ZZ \}$ as:  
\begin{equation} \label{quality-level-decision-wiflix}
m_u(t) = \mbox{argmin} \left \{Q_u(t)B_{f_u}(m,t) - \Theta_u(t) D_{f_u}(m,t) \; : \; m \in \{1, \ldots, N_{f_u}\} \right \}.
\end{equation}
As defined in (\ref{zero-def-wiflix}), for all transmission slots $t$ which are not integer multiples of $n$, there is no chunk requested and therefore $B_{f_u}(m_u(t),t)$ and $D_{f_u}(m_u(t),t)$ are equal to be $0$.
The second term in (\ref{DPP-wiflix}), after a change of sign, is given by 
\begin{align}
\sum_{u \in \Uc} \left \{ V \phi_u(\gamma_u(t)) -  \gamma_u(t) \Theta_u(t) \right \}. 
\end{align}
Again, this is maximized by maximizing separately each term, yielding the simple one-dimensional maximization (e.g., solvable by line-search):
\begin{equation} \label{opt-gamma-wiflix}
\gamma_u(t) = \mbox{argmax} \left \{  V\phi_u(\gamma) - \Theta_u(t) \gamma \; : \;  \gamma \in [D_u^{\min}, D_u^{\max}] \right \},
\end{equation}

We refer to the policy (\ref{quality-level-decision-wiflix}) as {\em pull congestion control} since 
each user $u$ selects the quality level at which this chunk is requested by taking into account the state of its request queue $Q_u$. It chooses an appropriate video quality level that balances the desire for high quality (reflected by the term $-\Theta_u(t) D_{f_u}(m,t)$ in (\ref{quality-level-decision-wiflix})) and the desire for low request queue lengths (reflected by the term  $ Q_u(t)B_{f_u}(m,t)$ in (\ref{quality-level-decision-wiflix})) and then opportunistically pulls the chunk at that video quality level from the helpers in its current vicinity. This policy is reminiscent of the current DASH technology \cite{sanchez2011idash}, where the client (user) progressively fetches a video file by downloading successive chunks, and makes adaptive decisions on the source encoding quality based on its current knowledge of the congestion of the underlying server-client connection.
Notice also that, in order to compute (\ref{quality-level-decision-wiflix}) and (\ref{opt-gamma-wiflix}), 
each user needs to know only {\em local information} formed by the locally maintained request queue backlog $Q_{u}(t)$ and by the locally computed 
virtual queue backlog $\Theta_u(t)$. 

\subsubsection{Control actions at the helper nodes (transmission scheduling)}
At transmission slot $t$, the network controller observes the queues $Q_u(t)$ of all users $u$ and the topology state $\omega(t)$, and chooses the feasible instantaneous rate matrix $[\mu_{hu}(t)] \in \Rc(\omega(t))$ to maximize the weighted sum rate of the transmission rates achievable in transmission slot $t$.
Namely, the network of helpers must solve the Max-Weighted Sum Rate (MWSR) problem:
\begin{align} \label{mwsr-general-wiflix}
\mbox{maximize} & \;\;\; \sum_{h \in \Hc} \sum_{u \in \Nc(h)} Q_{u}(t) \mu_{hu}(t) \nonumber \\
\mbox{subject to} & \;\;\; [\mu_{hu}(t)] \in \Rc(\omega(t)) 
\end{align}
where $\Rc(\omega(t))$ is the feasible instantaneous rate region of the network at slot $t$. 
It is immediate to see that, after a change of sign, the maximization of the third term in (\ref{DPP-wiflix}) yields the problem (\ref{mwsr-general-wiflix}).


%
\section{Policy Performance}
As outlined in Section~\ref{sec-sysmodel-wiflix}, VBR video yields time-varying quality and rate functions $D_f(m, t)$ and $B_f(m, t)$, which depend on the individual video file. Furthermore, arbitrary user motion yields slower time variations of the pathloss coefficients at the same time-scale of the video streaming session. As a result, any stationarity or ergodicity assumption about the topology state $\omega(t)$, the rate function $B_f(m, t)$ and quality function $D_f(m, t)$ is unlikely to hold in most practically relevant settings. Therefore, we consider the optimality of the DPP policy for an {\em arbitrary sample path} of the topology state $\omega(t)$, the quality function $D_f(m, t)$ and the rate function $B_f(m, t)$. Following in the footsteps of \cite{neely2010stochastic,neely2010universal}, we compare the network utility achieved by our DPP policy  with that achieved by an optimal oracle policy with  $T$-slot lookahead, i.e., knowledge of the future sample path over an interval of length $T$ slots. Time is split into frames of duration $T$ slots and we consider $F$ such frames.  For an arbitrary sample path of $\omega(t)$, $D_f(m, t)$ and $B_f(m, t)$, we consider the static optimization problem over the $j$-th frame
\begin{eqnarray}
\mbox{maximize} & & \sum_{u \in \Uc} \phi_u\left(\frac{1}{T}\sum_{\tau=jT}^{(j+1)T-1} D_{f_u}(m_{u}\left(\tau\right),\tau)\right)   \label{obj-arbit-wiflix}\\
\mbox{subject to} & & \frac{1}{T} \sum_{\tau=jT}^{(j+1)T-1}\left[ B_{f_u}\left(m_u(\tau),\tau\right)-\mu_{u}\left(\tau\right)\right]\leq0~\forall~u \in \Uc \label{arbit-const-wiflix} \\
& & [\mu_{hu}(\tau)] \in \Rc(\omega(\tau))~\forall~\tau~\in~\{jT, \ldots, (j+1)T-1\}, \label{feas-action-arbit-wiflix} \\
& & m_u(\tau) \in \{1, 2, \ldots, N_{f_u}\}~\forall~u \in \Uc,~\forall~\tau~\in~\{jT, \ldots, (j+1)T-1\}, \label{feas-qual-arbit-wiflix}
\end{eqnarray}
and denote by $\phi_j^{\rm opt}$ the maximum of the network utility function for 
frame $j$,  achieved over all policies which have future knowledge of the sample 
path over the $j$-th frame subject to the constraints (\ref{arbit-const-wiflix})-(\ref{feas-qual-arbit-wiflix}). We have the following result:
\begin{thm} \label{main-result-wiflix}
The DPP scheduling policy achieves per-sample path network utility
\begin{align}
\sum_{u \in \Uc} \phi_u \left( \overline{D}_u \right) \geq 
\lim_{F \rightarrow \infty} \frac{1}{F}\sum_{j=0}^{F-1}\phi_j^{\rm opt} -  O\left (\frac{1}{V} \right ) \label{optutil-arbit1-wiflix}
\end{align}
with bounded queue backlogs satisfying
\begin{align}
\lim_{F \rightarrow \infty} \frac{1}{FT} \sum_{\tau=0}^{FT-1}\left( \sum_{u \in \Uc}Q_{u}(\tau) 
+ \sum_{u \in \Uc}\Theta_{u}(\tau)\right) \leq  O(V) \label{strongstab-arbit1-wiflix}
\end{align}
where $O(1/V)$ indicates a term that vanishes as $1/V$ and $O(V)$ indicates a term that grows linearly with $V$, as the policy control parameter 
$V$ grows large. 
\end{thm}
\begin{IEEEproof} See Appendix \ref{proof-thm-wiflix}. \end{IEEEproof}

An immediate corollary of Theorem \ref{main-result-wiflix} is:
\begin{cor} \label{corcor-wiflix}
For the system defined in Section \ref{sec-sysmodel-wiflix}, when the evolution of the topology state $\omega(t)$, the rate function $B_f(m, t)$ and the quality function $D_f(m, t)$ is stationary and ergodic, then 
\begin{align}
\sum_{u \in \Uc} \phi_u(\overline{D}_u) \geq \phi^{\rm opt} - O\left ( \frac{1}{V}\right ), 
\label{utilperf-iid-wiflix}
\end{align}
where $\phi^{\rm opt}$ is the optimal value of the NUM problem (\ref{NUMproblem-wiflix}) in the stationary ergodic 
case,\footnote{Notice that in the stationary and ergodic case the value $\phi^{\rm opt}$ is generally achieved
by an instantaneous policy with perfect knowledge of the state statistics or, equivalently, by a policy with infinite look-ahead, since
the state statistics can be learned arbitrarily well from any sample path with probability 1, because of ergodicity.} 
and
\begin{align}
\sum_{u \in \Uc} \overline{Q}_{u} + \sum_{u \in \Uc}\overline{\Theta}_u \leq  O(V).
\label{strongstab-iid-wiflix}
\end{align}
In particular, if the network state is i.i.d., the bounding term in (\ref{utilperf-iid-wiflix}) is explicitly given by 
$O(1/V) = \frac{\Kc}{V}$, and the bounding term in (\ref{strongstab-iid-wiflix}) is explicitly given by 
$\frac{\Kc + V( \phi_{\max} - \phi_{\min} )}{\epsilon}$, where 
$\phi_{\min} = \sum_{u \in \Uc}\phi_u(D_u^{\min})$, $\phi_{\max} = \sum_{u \in \Uc}\phi_u(D_u^{\max})$, 
$\epsilon > 0$ is the slack variable corresponding to the constraint (\ref{arbit-const-wiflix}), 
and the constant $\Kc$ is defined in (\ref{dpp-ineq-wiflix}). 
\end{cor}

\begin{IEEEproof} See Appendix \ref{proof-thm-wiflix}. \end{IEEEproof}






\section{Wireless System Model with Massive MU-MIMO Helpers}\label{sec-channel-model-wiflix}

In this section, we first specify the region of instantaneous service rates $\Rc(\omega(t))$ for the specific PHY layer model comprising of massive MU-MIMO at each helper. We then
specialize the weighted sum-rate maximization problem (\ref{mwsr-general-wiflix}) to this system. By exploiting the channel-hardening effect of high dimensional MIMO channels, we observe that the MWSR problem is optimally solved by a low complexity greedy algorithm which can be implemented in a distributed manner with each helper independently choosing user subsets for MU-MIMO beamforming. 
\subsection{Helpers with Massive MU-MIMO}
Each helper $h$, with a large number of antennas $M$ installed, implements MU-MIMO to serve the users $\Nc(h)$ in its vicinity. As a result, helper $h$ can serve simultaneously, in the spatial domain, any subset of size not larger than $\min\{M,|\Nc(h)|\}$ of the users in $\Nc(h)$. We further assume that each helper performs linear zero-forcing beamforming (LZFBF) to the set of selected users (referred to in the following as ``active users"). 

The wireless channel is modeled by the well-known and widely accepted block-fading model, where at each transmission slot $t$, the channel corresponding to the helper-user link $(h,u)$ in $\Ec$ is given by 
\begin{equation}
y_u(t) = \sqrt{g_{hu}(t)}\xiv_{h,u}^\herm(t) \Vm_h(t) \xv_h(t) + \sum_{h' \neq h} \sqrt{g_{h'u}(t)}\xiv_{h',u}^\herm(t) \Vm_{h'}(t) \xv_{h'}(t) + z_u(t)
\end{equation}
where $\xiv_{h,u}(t)$ is the $M \times 1$ column vector of channel coefficients from the antenna array of helper $h$ to the receiving antenna of user $u$,  $g_{hu}(t)$ is the large-scale distance dependent pathloss from helper $h$ to user $u$, $\Vm_h(t)$ is the downlink precoding matrix of helper $h$, and $\xv_h(t)$ is the vector of transmitted complex information symbols (QAM modulation) of helper $h$.  $z_u(t)$ denotes the additive Gaussian noise at the $u$-th user receiver. Notice that this model takes fully into account the inter-cell interference of the signals sent by other helpers $h' \neq h$, on the link from helper $h$ to user $u$.

We use $\Sc_h(t)$ to denote the subset that is chosen for LZFBF in transmission slot $t$. The $M\times 1$ channel vectors $\xiv_{u,h}(t)$ of all users $u \in \Sc_h(t)$  are assumed to be known at the helper $h$ through some form of channel state feedback. Such channel vectors are collected as the columns of a $M\times |\Sc_h(t)|$ channel matrix $\Xim_h(t)$. The LZFBF precoded signal vector is given by $\Vm_h(t)\xv_h(t)$ where $\xv_h(t)$ is the $|\Sc_h(t)| \times 1$ column vector of symbols to be sent to users $u \in \Sc_h(t)$ and $\Vm_h(t)$ is the ZFBF precoding matrix of dimension $M  \times |\Sc_h(t)|$ given by the normalized pseudo-inverse
\begin{align}
\Vm(t) = \Xim_h(t)(\Xim_h^\herm(t)\Xim_h(t))^{-1}\Lambda(t)^{1/2}
\end{align}  
where $\Lambda(t)$ is a column-normalizing diagonal matrix with the $u$-th diagonal element given by
\begin{align}
\Lambda_u(t)=\frac{1}{\left[ \left(\Xim_h^\herm(t)\Xim_h(t)\right)^{-1}\right]_{uu}}
\label{wishart-wiflix}
\end{align}
where $[\cdot]_{uu}$ denotes the $u$-th diagonal element of the matrix argument. Using the fact that 
$\Xim_h^\herm(t)\Vm_h(t) = \Lambda(t)^{1/2}$, the resulting downlink channel to user $u \in \Sc_h(t)$  becomes
\begin{align}
y_u(t) = \sqrt{g_{hu}(t)\Lambda_u(t)}x_{hu}(t)+z_u(t)
\end{align}
where $g_{hu}(t)$ is the large scale pathloss coefficient from helper $h$ to user $u$. Under the assumptions that $M, |\Sc_h(t)| \rightarrow \infty$ with a fixed ratio$\frac{|\Sc_h(t)|}{M} \leq 1$, random matrix theory results (see \cite{bethanabhotla2014optimal, huh2012network}) can be invoked to show that 
\begin{align}
\Lambda_u(t) \rightarrow \left( 1- \frac{|\Sc_h(t)|-1}{M}\right)
\label{deterministic-wiflix}
\end{align}
Thus, for a given choice of subset $\Sc_h(t)$ and under the assumption that the power $P_h(t)$ is equally shared across the user streams in $\Sc_h(t)$, the vector $\cv_h(\Sc_h(t),t) = \{ c_{hu}(\Sc_h(t),t)\}_{u \in \Uc}$ of rates (in bits per channel symbol) achieved by all the users in $\Nc(h)$ is given by\footnote{This rate expressions neglects the effect of {\em pilot contamination}, which arises in 
massive MIMO with TDD and open-loop channel estimation based on uplink pilots and channel reciprocity. While in the regime of 
infinite number of base station antennas and finite number of users, pilot contamination dominates the massive MIMO performance in a multi-cell network 
\cite{Marzetta-TWC10}, it is well-known that in the more-realistic regime of large but finite number of antennas this effect is typically negligible with respect to
the residual multiuser inter-cell interference \cite{Huh11, hoydis2011massive}.
Here, for simplicity of exposition and space limitation, we neglect these effects and assume that the LZFBF precoder is computed from ideal knowledge of the channel matrix, 
such that our rate expressions are {\em exact} under this assumption. However, we hasten to say that our approach is immediately applicable to the case of imperfect channel state knowledge, 
by using the appropriate (more involved) feasible rate expressions.}
such that 
\begin{align}
 c_{hu}(\Sc_h(t),t) = \twopartdef {0} {u \notin \Sc_h(t)} {\log\left(1+\frac{M-|\Sc_h(t)|+1}{|\Sc_h(t)|}\frac{P_hg_{hu}(t)}{1+\sum_{h' \neq h}P_{h'u}g_{h'u}(t)}\right)} {u \in \Sc_h(t)}
\label{equal power-wiflix}
\end{align}
In fact, it is known that the asymptotics kick in very quickly making the rates in~(\ref{equal power-wiflix}) achievable for practical values of $M$ and $|\Sc_h(t)|$. Notice that the rate expression is independent of the small scale fading coefficients. This is because of using a large number of antennas $M$ at the helpers which renders a large $M \times |\Sc_h(t)|$ random channel matrix $\Xim(t)$ of i.i.d complex Gaussian small scale fading coefficients in every transmission slot $t$. When each helper performs LZFBF in every transmission slot $t$, the coefficients $\Lambda_u(t)$ given in (\ref{wishart-wiflix}) by the reciprocals of the diagonal elements of the inverse Wishart matrix $(\Xim^\mathrm{H}(t)\Xim(t))^{-1}$ ``harden'' at a deterministic value (\ref{deterministic-wiflix}) (see \cite{huh2012network}) due to the large size of the matrix $\Xim(t)$ and the assumption $\frac{|\Sc_h(t)|}{M} \leq 1$. This results in deterministic rate expressions as in (\ref{equal power-wiflix}) which are independent of $\Xim(t)$ and are just dependent on the large scale path loss coefficients $g_{hu}(t)$.
Furthermore, in the case when the helpers are incapable of MU-MIMO, i.e., when the active user subset size $|\Sc_h(t)|$ is chosen to be exactly $1$, the above formula still holds by setting $|\Sc_h(t)|=1$ and this is referred to as single user MIMO (SU-MIMO).

 Since helper $h$ can choose an active user subset from the collection of all possible user subsets of $\Nc(h)$, the vector $\{\mu_{hu}(t)\}_{u \in \Nc(h)}$ of bits scheduled by helper $h$ to users $u$ in its neighborhood $\Nc(h)$ is constrained to lie in the discrete set of vectors
\begin{align}
\{s\cv_h(\Sc_h,t): \Sc_h \subseteq \Nc(h)\}
\label{massive-polytope-wiflix}
\end{align}
where $s$ is the number of channel symbols available in every transmission slot $t$. Notice from the rate expression in (\ref{equal power-wiflix}) that the topology state $\omega(t)$ in this wireless system is given by the vector $\{g_{hu}(t)\}$ of large-scale pathloss coefficients between each helper-user pair $(h, u) \in \Ec$. 

We assume that the receiver at every user is {\it advanced} in the sense that it can decode multiple streams in the same transmission slot, i.e., user $u$, in transmission slot $t$, can receive  $\mu_{u}(t)= \sum_{h \in \Nc(u)}\mu_{hu}(t)$ video-encoded bits by simultaneously downloading $\mu_{hu}(t)$ bits from helpers $h$ in $\Nc(u)$. Notice that each stream is achievable (in an information theoretic sense), 
by treating the other streams as Gaussian  noise,  i.e., we do not make use of multiuser detection schemes (e.g., based on successive interference cancellation) at the user receivers. Therefore, our rate expressions 
are representative of what can be achieved with today's user device technology. 

For the sake of comparison, in the simulation results of  Section \ref{sec-simul-wiflix} we also consider a {\it dumb receiver heuristic} 
where each user $u$ decodes only the strongest data stream and therefore downloads only $ \max_{h\in \Nc(u)}\mu_{hu}(t)$ video-encoded bits. 
While the dumb receiver heuristic is a degradation of the optimal solution involving advanced receivers,  the simulation results in Section \ref{sec-simul-wiflix} show that this degradation is almost negligible. 
This also implicitly indicates that, in most relevant practical topologies and pathloss scenarios, it is unlikely that the same user is scheduled by more than one helper in the same transmission slot.
 
\subsection{Transmission Scheduling with Massive MU-MIMO Helpers}\label{cor-mimo-proof-wiflix}

We now particularize the problem (\ref{mwsr-general-wiflix}) to the specific wireless system with massive MU-MIMO helpers. For the constraint (\ref{massive-polytope-wiflix}) specific to the wireless system, the general weighted sum-rate maximization problem (\ref{mwsr-general-wiflix}) reduces to:
\begin{eqnarray}
\mbox{maximize} & & \sum_{h \in \Hc} \sum_{u \in \Nc(h)} Q_{u}(t) \mu_{hu}(t) \nonumber \\
\mbox{subject to} & & \{\mu_{hu}(t)\}_{u \in \Nc(h)} \in \{s\cv_h(\Sc_h,t): \Sc_h \subseteq \Nc(h)\}~\forall~h \in \Hc.
\end{eqnarray}
 This problem decouples into separate maximizations for each helper $h$ given by the following discrete optimization problem:
\begin{eqnarray}
\mbox{maximize} & & \sum_{u \in \Nc(h)} Q_{u}(t) \mu_{hu}(t) \nonumber \\
\mbox{subject to} & & \{\mu_{hu}(t)\}_{u \in \Nc(h)} \in \{s\cv_h(\Sc_h,t): \Sc_h \subseteq \Nc(h)\}. \label{dist-helper-prob-wiflix}
\end{eqnarray} 
The above optimization problem at each helper $h$ essentially corresponds to maximizing the weighted sum rate over the discrete set of vectors $\{s\cv_h(\Sc_h,t): \Sc_h \subseteq \Nc(h)\}$ with an exponential number $2^{|\Nc(h)|}-1$ of choices for the active user subset. However, the key observation from rate expression (\ref{equal power-wiflix}) is that when helper $h$ schedules the subset $\Sc_h$ of users for MU-MIMO beamforming, the rate of each user $u \in \Sc_h$ depends only on the cardinality $|\Sc_h|$ but not on the identity of the members of the subset $\Sc_h$. This implies that for a fixed subset size $S$, the subset $\Uc^*(S,t)$ of users maximizing the weighted sum rate can be obtained by sorting the users in $\Nc(h)$ according to the weighted rate $Q_u(t) \log\left(1+\frac{M-S+1}{S}\frac{P_hg_{hu}(t)}{\left(1+\sum_{h' \neq h}P_{h'u}g_{h'u}(t)\right)}\right)$ and choosing greedily the best $S$ users. Thus, we have
\begin{equation}  \label{sort-users-wiflix}
\Uc^*_h(S,t) = \mbox{argmax-}S \left \{ Q_{u}(t) \log\left(1+\frac{M-S+1}{S}\frac{P_hg_{hu}(t)}{1+\sum_{h' \neq h}P_{h'u}g_{h'u}(t)}\right) \; : \; u \in \Nc(h) \right \}, 
\end{equation}
where $\mbox{argmax-}S$ denotes the operation of choosing the first $S$ elements of a set of real numbers sorted in decreasing order.

This {\em sort \& greedy selection} procedure is repeated for every subset size yielding all the subsets $\{ \Uc^*(S,t) \}_{S = 1}^{|\Nc(h)|}$. Then, from these subsets, the subset $\Uc^*(t)$ which has the maximum weighted sum rate is picked as
\begin{equation}\label{subset-selection-wiflix}
\Uc^*_h(t) = \mbox{argmax} \left \{ \sum_{u \in \Uc^*_h(S,t)} Q_{u}(t) \log\left(1+\frac{M-S+1}{S}\frac{P_hg_{hu}(t)}{1+\sum_{h' \neq h}P_{h'u}g_{h'u}(t)}\right)\; : \; \Uc^*_h(S,t)~\forall~S \right \}
\end{equation}
 yielding the optimal solution to (\ref{dist-helper-prob-wiflix}). 

A typical sorting algorithm has complexity $O\left(|\Nc(h)|\log(|\Nc(h)|)\right)$ and since the sorting procedure is repeated for every subset size, the algorithm has complexity $O\left(|\Nc(h)|^2\log(|\Nc(h)|\right)$ which improves upon existing user scheduling algorithms\cite{yoo2006optimality} for the MIMO broadcast channel.

\section{Pre-buffering and re-buffering chunks}  \label{sec-prebuffering-wiflix} 
As described in Section \ref{sec-sysmodel-wiflix}, the playback process consumes chunks at a fixed playback rate $1/T_{\mathrm{gop}}$ (one chunk per video chunk slot $i$), while the number of chunks received per video chunk slot is a random variable due to the fact that $\omega(t)$ is a random process and the transmission resources are dynamically allocated by the DPP scheduling policy. In order to prevent stall events, each user $u$ should choose its pre-buffering time $T_u$ to be larger than the maximum delay with which a chunk is delivered to it. However, such maximum delay is neither deterministic nor known a priori. Moreover, even in special cases where the maximum delay of each request queue in the system admits a deterministic bound (e.g., see \cite{neely2012wireless}), such a bound may be loose and setting the pre-buffering time to be larger than that bound might be simply unacceptable in a practical system implementation. We therefore follow the scheme in \cite{bethanabhotla2013joint} where each user $u$ estimates its local delays by monitoring its delivery times in a sliding window spanning a fixed number of video chunk slots. However, the key difference from \cite{bethanabhotla2013joint} is that the scheme in this paper is much simpler since the proposed pull congestion control scheme ensures that chunks are received in the right playback order.

The goal here is to determine the delay $T_u$ after which user $u$ should start playback, with respect to the time at which the first chunk is requested (beginning of the streaming session).  
We define the size of the playback buffer $\Psi_u(i)$ as the number of chunks available in the buffer at video chunk slot $i$ but not yet played out. Without loss of generality, 
assume that the streaming session starts at $i = 1$. Then, $\Psi_u(i)$ evolves at the video timescale over video chunk slots $i \in \{1, 2, 3, \ldots \}$ as:\footnote{$1\{\Kc\}$ denotes the indicator function
of a condition or event $\Kc$.}
\begin{align}
\Psi_u(i) = \max \left \{ \Psi_u(i-1)  -  1\{i > T_u\}, 0 \right \} + a_i.
\end{align}
where $a_i$ is the number of chunks which are completely downloaded in the transmission slots between $t=(i-1)n$ and $t=in$. Note that the playback buffer is updated every video chunk slot $i$, i.e., at the time scale of seconds. Thus, if the download of a chunk is completed between $t=(i-1)n$ and $t=in$, from the playback buffer's perspective, the chunk is considered to have arrived at the end of the $i$-th video chunk slot, i.e., at $t=in$.
Let $A_k$ denote the video chunk slot in which chunk $k$ arrives at the user and let $W_k$ denote the delay (measured in video chunk slots) with which chunk $k$ is delivered. Note that the longest period during which $\Psi_u(i)$ is not incremented is given by the maximum delay to deliver chunks. Thus, each user $u$ needs to adaptively estimate $W_k$ in order to choose $T_u$. In the proposed method, at each video chunk slot $i= 1,2,\ldots$, user $u$ calculates the maximum observed delay $E_i$  in a sliding window of size $\Delta$, by letting:
\begin{align}
E_i = \max \{ W_k \; :  ~i-\Delta+1 \leq A_k \leq i \}.
\label{del-window-wiflix}
\end{align}
Finally, user $u$ starts its playback when $\Psi_i$ crosses the level $\rho E_i$, i.e., $T_u = \min \{ i : ~\Psi_u(i) \geq \rho E_i \}$ where $\rho$ is an algorithm control parameter. If a stall event occurs at video chunk slot $T$, i.e., $\Psi_i = 0$ for $i > T$, the algorithm enters a re-buffering phase in which the same algorithm presented above is employed again to determine the new instant $T+ T_u + 1$ at which playback is restarted. With slight abuse of notation, we have re-used $T_u$ to denote the re-buffering delay although it is re-estimated using the sliding window method at each new stall event.

\section{Numerical Experiment} \label{sec-simul-wiflix}
                                                            
\begin{figure}
\centering
\includegraphics[scale=0.35]{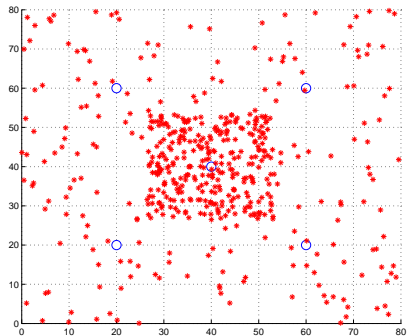}
\caption{Simulation setup}
\label{topology-wiflix}
\end{figure}

Our simulations are based on a network topology formed by a 80m$\times$80m region with $5$ helpers (indicated by $\circ$'s) as shown in Fig.~\ref{topology-wiflix}. The users (indicated by $*$'s)  are generated according to a non-homogeneous Poisson point process with higher density in a central region of size $\frac{80}{3}$m$\times$$\frac{80}{3}$m, as shown in Fig.~\ref{topology-wiflix}. 

Each helper has $M$ antennas and serves user sets of size upto $S$, with transmission power of $35$dBm. The pathloss from a helper to a user is given by $\frac{1}{1+(\frac{d}{40})^{3.5}}$, with $d$ representing the helper-user distance (assuming a torus wrap-around model to avoid boundary effects).  
We assume a PHY fame duration of $10$ ms and a total system bandwidth of $18$ Mhz as specified in the LTE 4G standard. With one OFDM resource block ($7 \times 12$ channel symbols) spanning  $0.5$ ms in time and $180$ Khz in bandwidth (corresponding to $12$ adjacent subcarriers each with $15$ KHz bandwidth), each transmission slot spans $s=84\times100\times20$ channel symbols.   

We assume that all the users request chunks successively from VBR-encoded video sequences.  
Each video file is a long sequence of chunks, each of duration $0.5$ seconds and with a frame rate $30$ frames per second. We consider a specific video sequence formed by $800$ chunks, constructed using several standard video clips from the database in~\cite{video-samples}. The chunks  are encoded into different quality modes with the quality index measured using the {\em Structural SIMilarity} (SSIM) index defined in \cite{ssim}. The chunks from $1$ to $200$ are encoded into $8$ quality modes with an average bitrate of $631$ kbps. Chunks $201$ to $400$ are encoded in $4$ quality modes at an average bitrate of $3908$ kbps. Similarly, chunks $401-600$ and $601-800$ are encoded into $4$ and $8$ quality modes with average bitrates of $6679$ kbps and $556$ kbps respectively. In the simulation, each user starts its streaming session of $1000$ chunks from some arbitrary position in this reference  video sequence and successively requests $1000$ chunks by cycling through the sequence.
 
 We choose the utility function $\Phi_u(\cdot) = \log(\cdot)~\forall~u \in \Uc$ to impose proportional fairness. We set the pre-buffering algorithm control parameter (described in Section \ref{sec-prebuffering-wiflix}) $\rho=3$. We simulate our algorithm for the layout shown in Fig.~\ref{topology-wiflix} (with around $500$ users generated according to a non-homogenous Poisson point process as explained above). At $t=1$, all the users simultaneously start streaming $1000$ chunks. 
  
 We studied the performance of our algorithm with $M=40$ antennas and maximum active user subset size $S=10$ for different values of the policy control parameter $V$
and observed that both the QoE metrics average video quality and the $\%$ of time spent in buffering mode are satisfactory for the choice of $V=2*10^{14}$. We use that value for the rest of the simulations in this section.
 
\begin{figure}
\subfloat[]{
\includegraphics[scale=0.28]{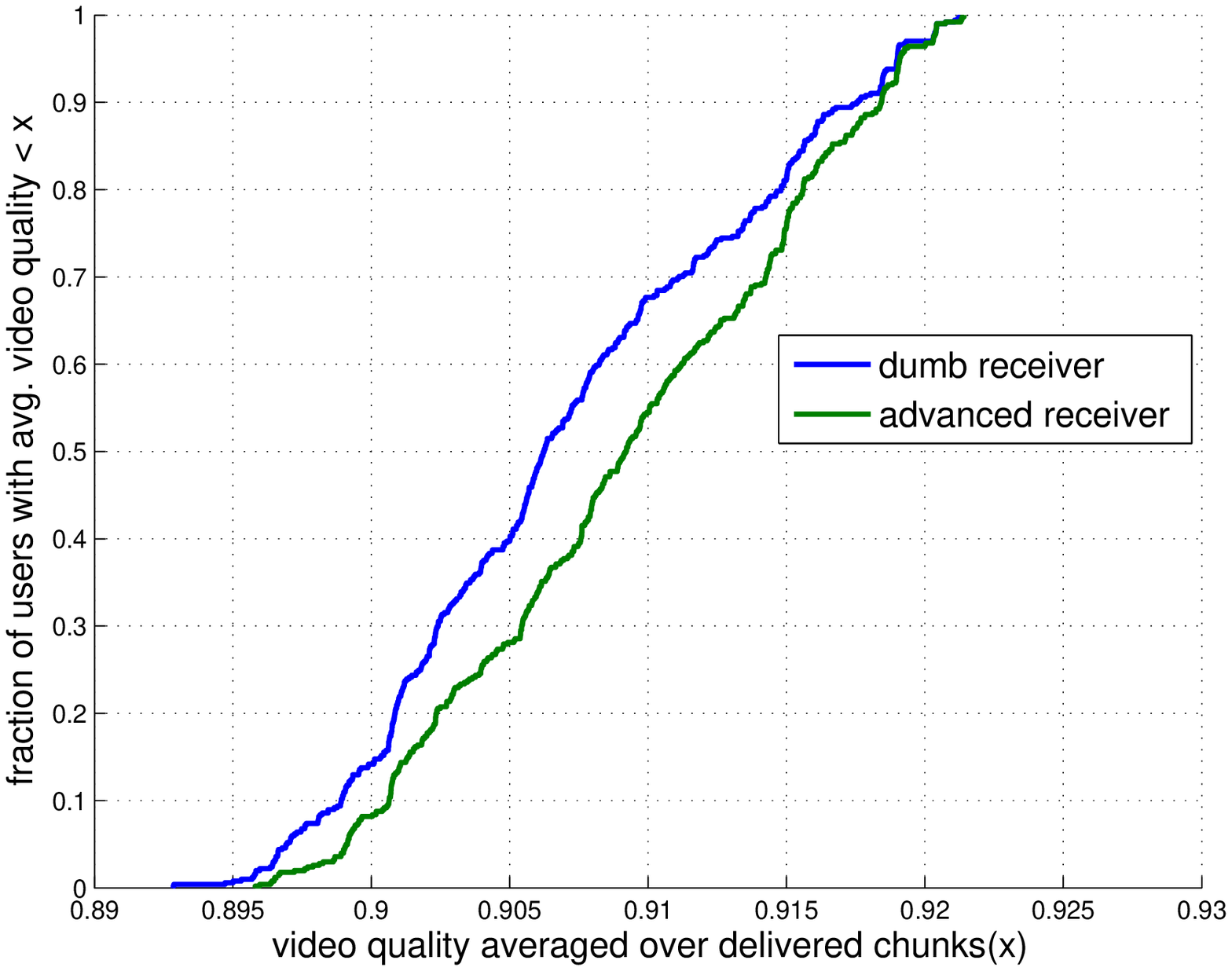}
\label{ssim-cdf-dumb-wiflix}
}
\subfloat[]{
\includegraphics[scale=0.28]{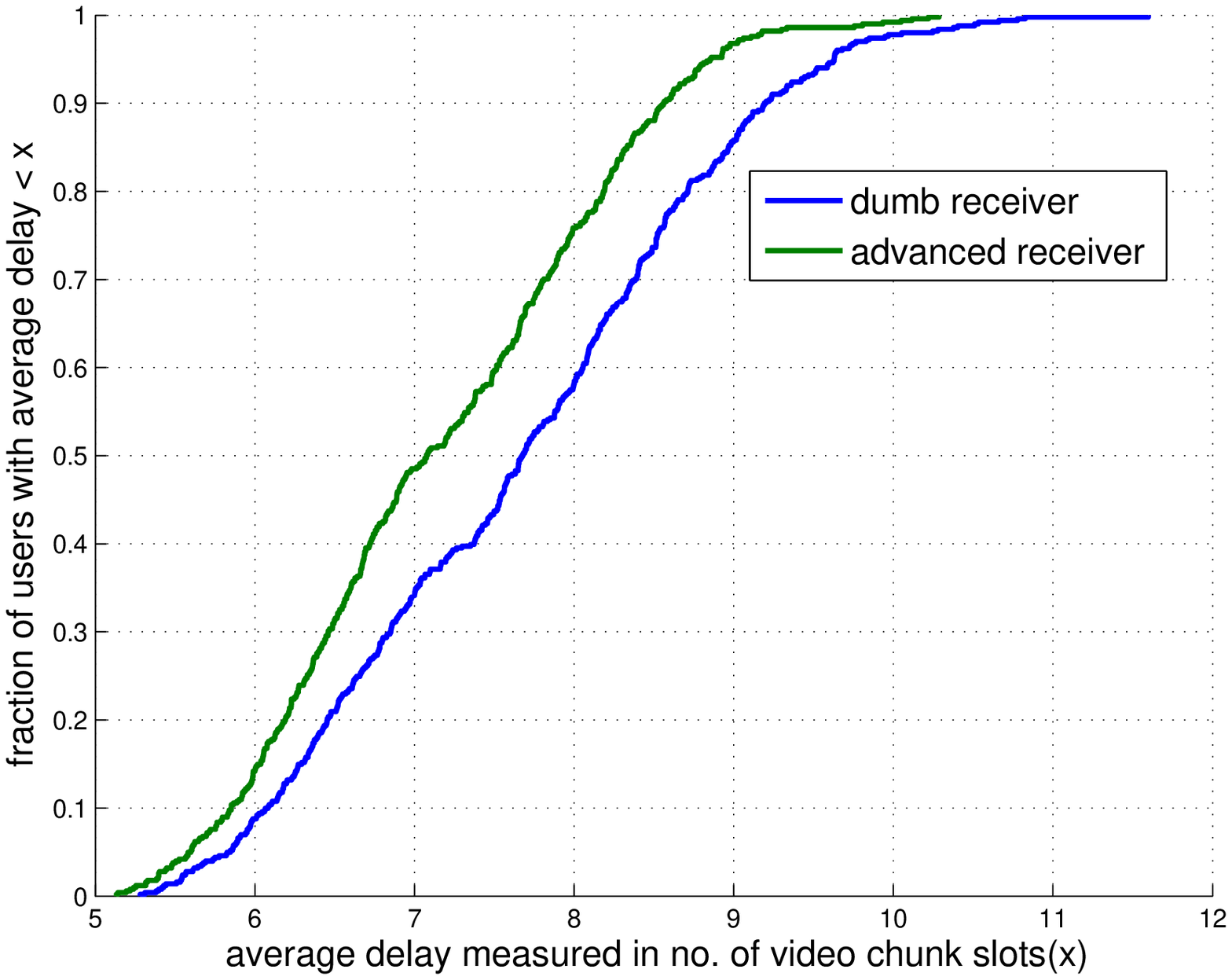}
\label{delay-cdf-dumb-wiflix}
}
\subfloat[]{
\includegraphics[scale=0.27]{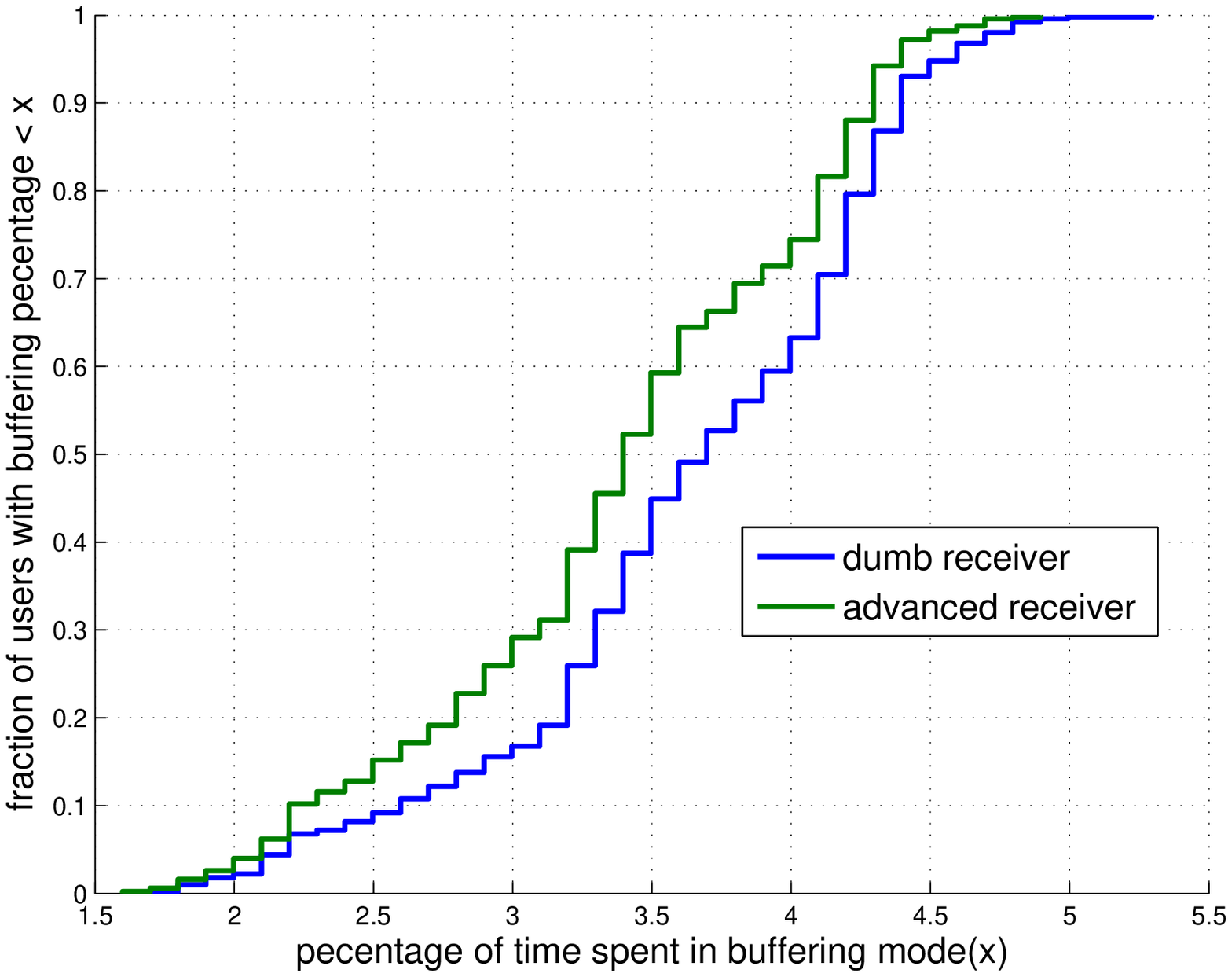}
\label{buff-cdf-dumb-wiflix}
}
\caption{Performance comparison of advanced and dumb receivers.}
\end{figure}

 We now study the performance loss experienced under the {\it dumb receiver heuristic} where the receiver at every user $u$ decodes only the strongest signal and downloads only $\max_{h \in \Nc(u)} \mu_{hu}(t)$ in contrast to the macro diversity advanced receiver which can decode multiple signals simultaneously and download all the $\sum_{h \in \Nc(u)}\mu_{hu}(t)$ bits. Using $M=40$ and $S=10$, we simulate our algorithm and plot the CDF's over the user population of a) the average video quality b) the average delay in the reception of video chunks measured in video chunk slots and c) the $\%$ of playback time spent in buffering mode in Figs.~\ref{ssim-cdf-dumb-wiflix}, \ref{delay-cdf-dumb-wiflix} and \ref{buff-cdf-dumb-wiflix} respectively. We observe that the performance loss in using a dumb receiver is fairly negligible and therefore use a dumb receiver for the rest of the simulations in this section.

\begin{figure}
\subfloat[]{
\includegraphics[scale=0.35]{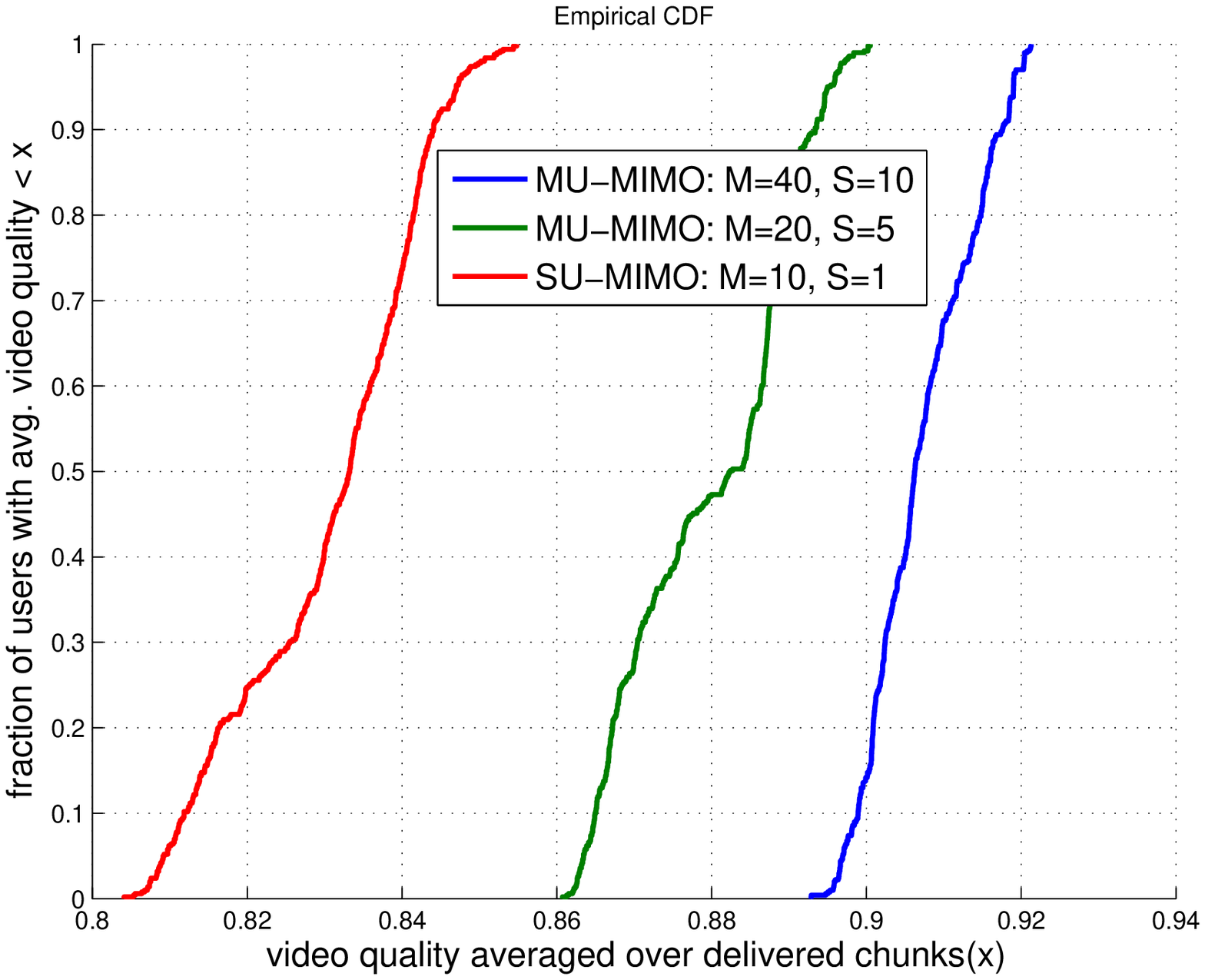}
\label{ssim-cdf-mimo-wiflix}
}
\subfloat[]{
\includegraphics[scale=0.35]{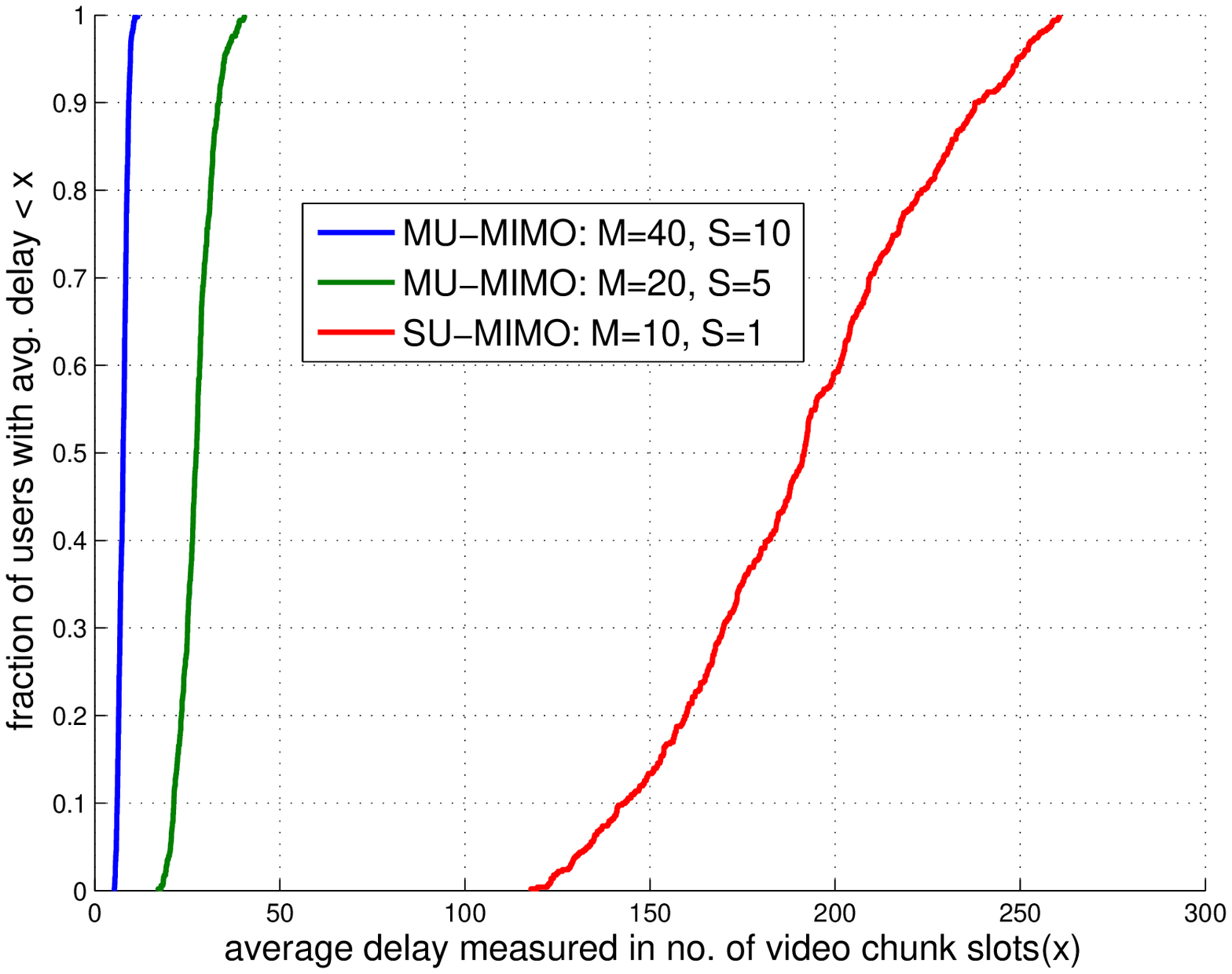}
\label{delay-cdf-mimo-wiflix}
}\\
\centering
\subfloat[]{
\includegraphics[scale=0.35]{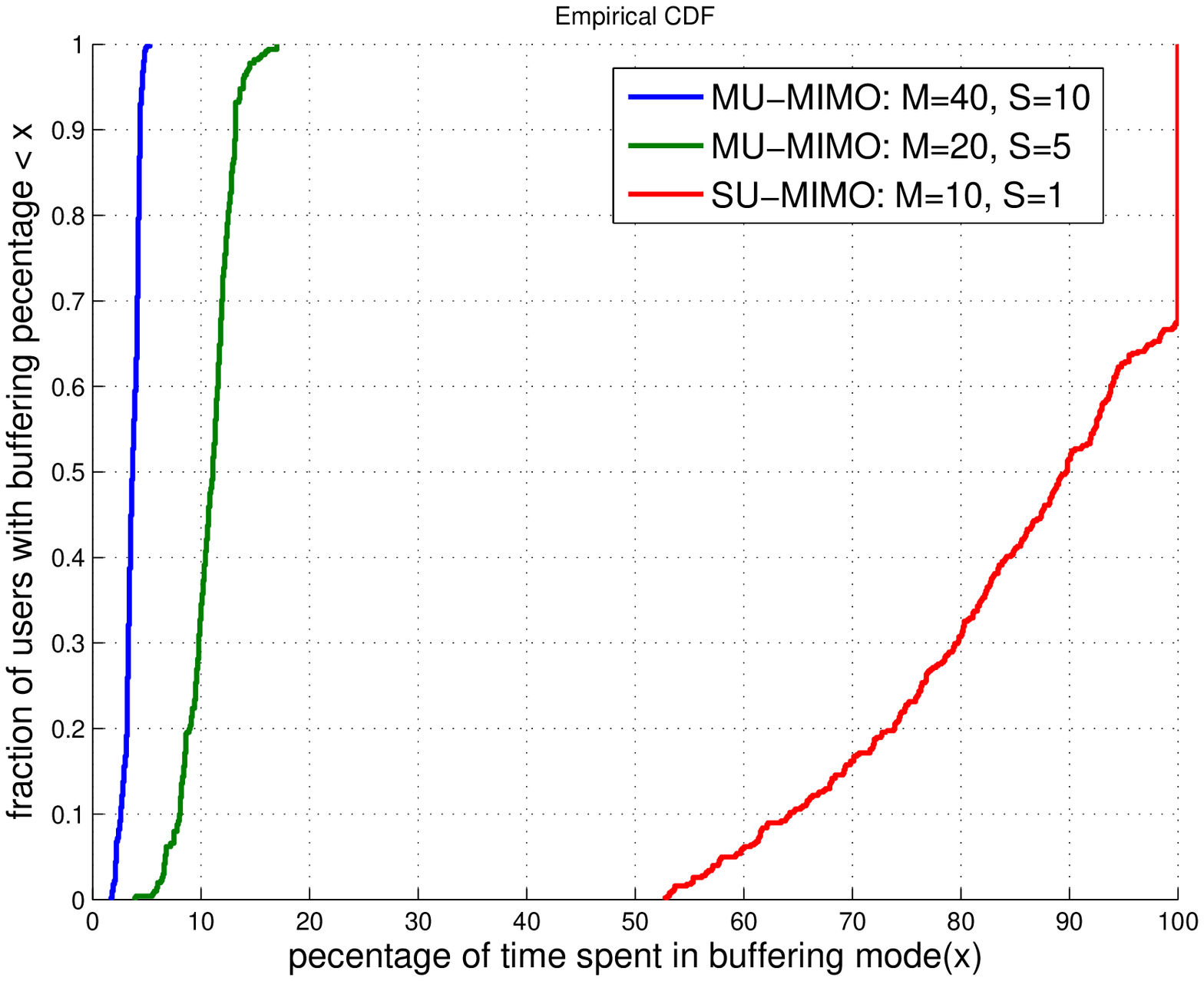}
\label{buff-cdf-mimo-wiflix}
}
\caption{Video streaming QoE improvement with MU-MIMO over SU-MIMO}
\end{figure}

We next study the QoE improvement when MU-MIMO is deployed at the helpers in place of legacy SU-MIMO systems. We plot the CDF over the user population of the same video streaming QoE metrics as in the previous figures for three different cases 1) MU-MIMO with $M=40$ antennas and maximum active user subset size $S=10$; 2) MU-MIMO with $M=20$ antennas and maximum active user subset size $S=5$; 3) SU-MIMO with $M=10$ antennas. From Figs. \ref{ssim-cdf-mimo-wiflix}, \ref{buff-cdf-mimo-wiflix} and \ref{delay-cdf-mimo-wiflix}, we can observe that there is significant improvement of video streaming performance in terms of the average video quality, the average delay (or alternately the percentage of time spent in buffering mode) when MU-MIMO is employed at the PHY layer in comparison to SU-MIMO. This clearly indicates that upgrading current SU-MIMO systems to massive MU-MIMO is a promising approach to meet the increasing demands for HD video streaming.

\begin{figure}
\centering
\subfloat[]{
\includegraphics[scale=0.35]{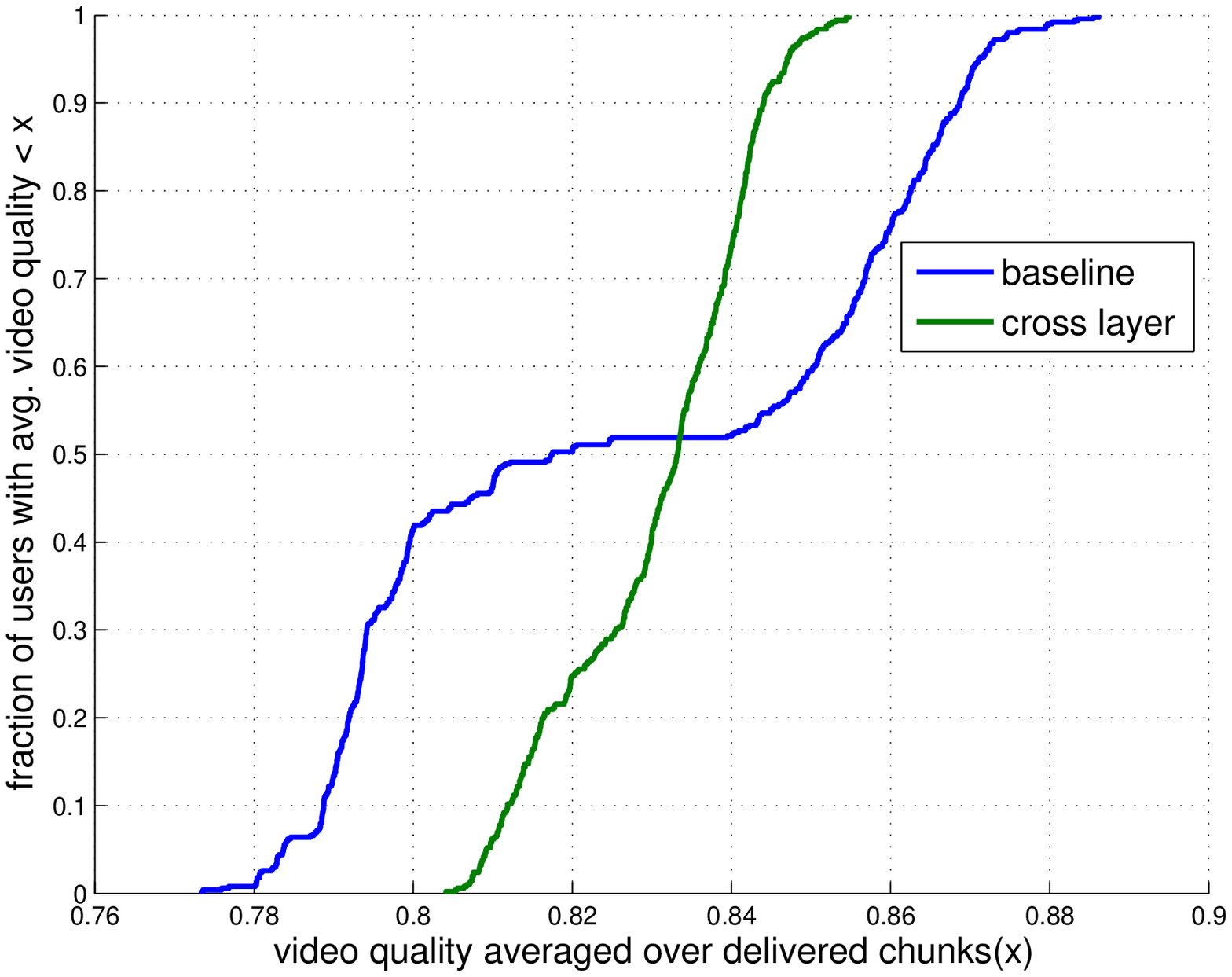}
\label{ssim-soa-wiflix}
}
\subfloat[]{
\includegraphics[scale=0.35]{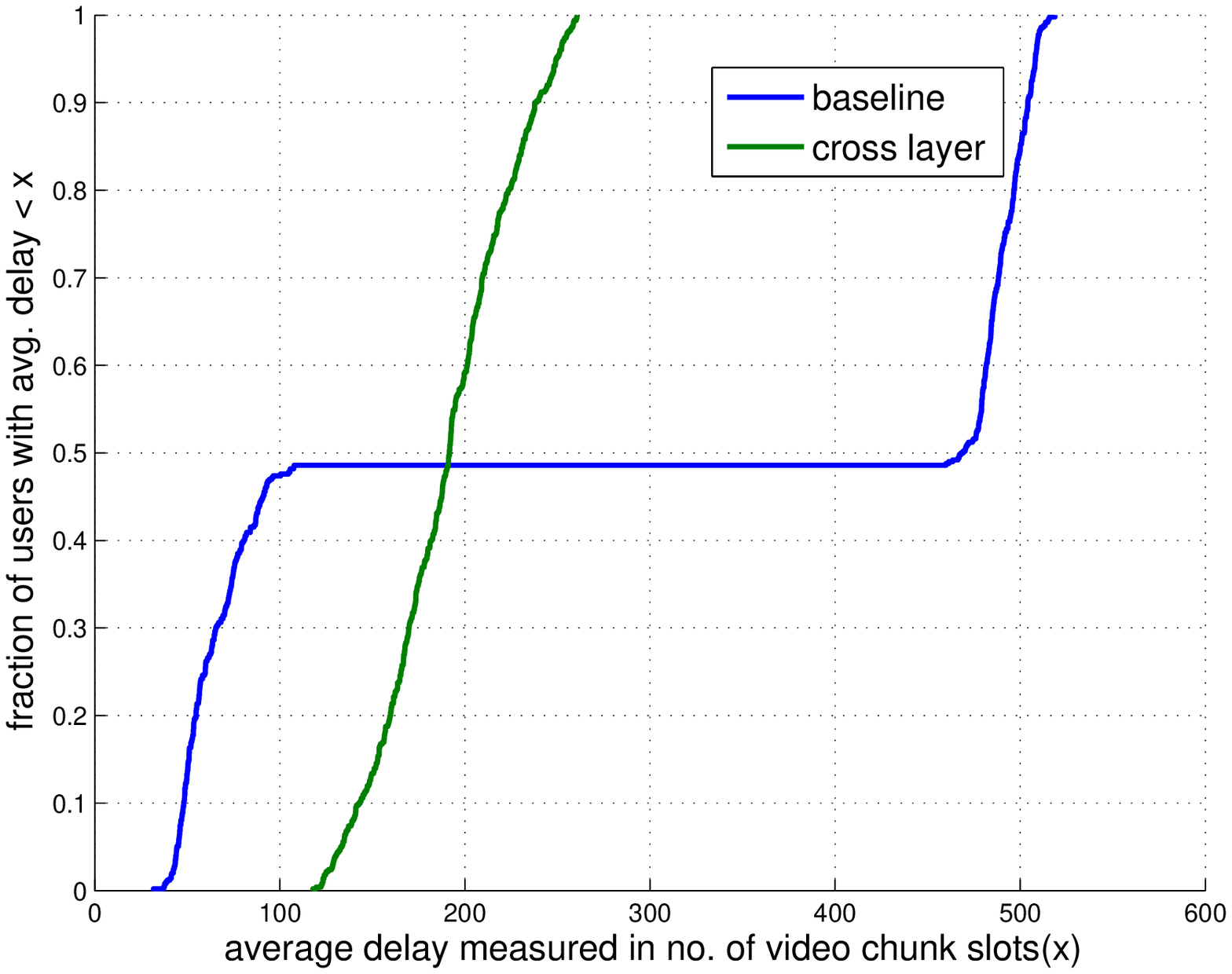}
\label{avgdelay-soa-wiflix}
}
\caption{Performance comparison of a cross-layer approach with a baseline scheme.}
\end{figure}

Finally, we study the benefits of using a cross layer approach in comparison to a baseline scheme representative of legacy wireless systems. We perform this comparison for the case where every helper employs SU-MIMO with $M=10$ antennas. For the baseline scheme, every user first fixes its association with the unique helper that provides the maximum received signal strength (RSSI) $P_hg_{hu}$ and then uses the same control decision (\ref{quality-level-decision-wiflix}) to choose the quality levels for the chunks that arrive into the request queue every video chunk slot. Furthermore, we assume that the helpers {\it locally} employ proportional fairness/ equal air-time scheduling, i.e.,  each helper $h$ schedules the users associated with it through the max-RSSI scheme in a round-robin fashion across the transmission slots {\it independent of the request queue lengths at the users}. This baseline scheme is representative of current practical systems where the decisions across different layers are independent and there is no interaction between the upper and lower layers. We plot the CDFs over the user population of the average video quality and the average delay in the reception of chunks in Figs.~\ref{ssim-soa-wiflix} and \ref{avgdelay-soa-wiflix} respectively. We can observe that the cross layer scheme treats the users in a fair manner while the baseline scheme favors some users at the expense of other users in the system.
\appendices

\section{Proof of Theorem \ref{main-result-wiflix} and of Corollary \ref{corcor-wiflix}}  \label{proof-thm-wiflix}

 As in Section~\ref{sec-formulation-wiflix}, we consider the following problem, equivalent to (\ref{obj-arbit-wiflix}) -- (\ref{feas-qual-arbit-wiflix}), which involves a sum of time-averages instead of functions of time averages and introduces the auxiliary variables $\gamma_u(t)$:
 {\small
\begin{eqnarray}
\label{constraint-wiflix} 
\textrm{maximize} & & \frac{1}{T}\sum_{\tau=jT}^{(j+1)T-1} \sum_{u \in \Uc}\phi_u\left(\gamma_u(\tau)\right)   \\
\textrm{subject to} & & \frac{1}{T} \sum_{\tau=jT}^{(j+1)T-1}\left[
B_{f_u}\left(m_u(\tau),\tau\right)-\mu_{u}\left(\tau\right)\right]\leq0~\forall~u \in \Uc \label{qstable-arbit-wiflix} \\
& & \frac{1}{T} \sum_{\tau=jT}^{(j+1)T-1}\left[
\gamma_u\left(\tau\right)-D_{f_u}\left(m_u(\tau), \tau\right)\right]\leq0~\forall~ u \in \Uc \label{virt-stable-arbit-wiflix}\\
& &  D_u^{\min} \leq \gamma_u(\tau) \leq D_u^{\max}~\forall~u \in \Uc,~\forall~\tau \in \{jT, \ldots, (j+1)T-1\} \label{gamma-limit-wiflix} \\
& & [\mu_{hu}(\tau)] \in \Rc(\omega(\tau))~\forall~\tau \in \{jT, \ldots, (j+1)T-1\} \label{feasactions-trans-arbit-wiflix}\\
& & m_u(\tau) \in \{1, 2, \ldots, N_{f_u}\}~\forall~u \in \Uc,~\forall~\tau \in \{jT, \ldots, (j+1)T-1\} \label{feasqual-arbit-wiflix}
\end{eqnarray}
}
The update equations for the request queues $Q_{u}~\forall~u \in \Uc$ and the virtual queues
$\Theta_u~\forall~u \in \Uc$ are given in (\ref{q-update-wiflix}) and in (\ref{virt-update-wiflix}), respectively. Let ${\bf G}(\tau)=\left[ {\bf Q}^\transp(\tau), \Thetav^\transp(\tau)\right]^\transp$ be the combined queue backlogs column vector, and define the quadratic Lyapunov function $L(\Gm(\tau) = \frac{1}{2} \Gm^\transp (\tau) \Gm(\tau)$. Fix a particular slot $\tau$ in the $j$-th frame. We first consider the one-slot drift of $L({\bf G}(\tau))$. From~(\ref{drift-bound-wiflix}), 
we know that
\begin{align}
L({\bf G}(\tau+1))-L({\bf G}(\tau)) &\leq  \Kc + \left({\bf B}(\tau)-{\boldsymbol \mu}(\tau)\right)^
\transp{\bf Q}(\tau) + \left(\gammav(\tau)-
\Dm(\tau)\right)^\transp\Thetav(\tau)\label{drift-arbit-wiflix}
\end{align}
where $\Kc$ is a uniform bound on the term 
$$\frac{1}{2}\left[\left(\Bm(\tau)-\muv(\tau)\right)^\transp\left(\Bm(\tau)-\muv(\tau)\right)+\\ \left(\gammav(\tau)-\Dm(\tau)\right)^\transp\left(\gammav(\tau)-\Dm(\tau)\right)\right],$$ which exists under the realistic assumption that the chunk sizes, the transmission rates and the video quality measures are upper bounded by some constants, independent of $\tau$. We choose $\Kc$ such that
\begin{equation}
\label{B-definition-wiflix}
 \Kc > 2{\boldsymbol \kappa}^\transp{\boldsymbol \kappa}
\end{equation}
where  $\kappav$ is a vector whose components are all equal to the same number $\kappa$ and this number is a uniform upper bound on the maximum possible magnitude of drift in any of the queues (both the request queues $Q_u$ and the virtual queues $\Theta_u$) in one slot. With the additional penalty term $-V \sum_{u \in \Uc}\phi_u(\gamma_u(\tau))$ added on both sides of (\ref{drift-arbit-wiflix}), we have the following DPP inequality:
 \begin{align}
 L({\bf G}(\tau+1))-L({\bf G}(\tau))-V \sum_{u \in \Uc}\phi_u(\gamma_u(\tau)) \leq  \Kc &+ \left({\bf B}(\tau)-{\boldsymbol \mu}(\tau)\right)^
\transp{\bf Q}(\tau) + \left(\gammav(\tau)-
\Dm(\tau)\right)^\transp\Thetav(\tau) \notag \\
&-V \sum_{u \in \Uc}\phi_u(\gamma_u(\tau))\label{driftpluspenalty-arbit-wiflix}
 \end{align}
Let the DPP policy which minimizes the right hand side of the {\it drift-plus-penalty} inequality (\ref{driftpluspenalty-arbit-wiflix}) comprise of the control actions $\{m_u(\tau)\}_{\tau=jT}^{(j+1)T-1}~\forall~u \in \Uc$, $\{\gammav(\tau)\}_{\tau=jT}^{(j+1)T-1}$ and $\{(\mu_{hu}(\tau))\}_{\tau=jT}^{(j+1)T-1}$. Since the DPP policy minimizes the expression on the RHS of (\ref{driftpluspenalty-arbit-wiflix}), any other policy comprising of the control actions $\{m_u^*(\tau)\}_{\tau=jT}^{(j+1)T-1}~\forall~u \in \Uc$, $\{\gammav^*(\tau)\}_{\tau=jT}^{(j+1)T-1}$ and $\{(\mu_{hu}^*(\tau))\}_{\tau=jT}^{(j+1)T-1}$ would give a larger value of the expression. We therefore have
\begin{align}
 L({\bf G}(\tau+1))-L({\bf G}(\tau))-V \sum_{u \in \Uc}\phi_u(\gamma_u(\tau)) 
 \leq \Kc &+\left({\bf B}^*(\tau)-\muv^*(\tau)\right)^\transp{\bf Q}(\tau) + \left(\gammav^*(\tau)-
\Dm^*(\tau)\right)^\transp\Thetav(\tau) \notag \\
&-V\sum_{u \in \Uc}\phi_u(\gamma_u^*(\tau)). \label{alpha-star-wiflix}
\end{align}
Further, we note that the maximum change in the queue length vectors $Q_{u}(\tau)$ and $\Theta_u(\tau)$ from one slot to the 
next is bounded by $\kappa$. Thus, we have for all $\tau \in \{jT, \ldots, (j+1)T-1\}$
{\small
\begin{align}
 |Q_{u}(\tau)- Q_{u}(jT)| &\leq (\tau-jT)\kappa~~\forall~u \in \Uc \label{bounding1-wiflix}\\
|\Theta_u(\tau)-\Theta_u(jT)| &\leq (\tau-jT)\kappa~~\forall~u \in \Uc \label{bounding2-wiflix}
\end{align}
}
Substituting the above inequalities in~(\ref{alpha-star-wiflix}), we have
{\small
\begin{align}
 L({\bf G}(\tau+1))-L({\bf G}(\tau))-V \sum_{u \in \Uc}\phi_u(\gamma_u(\tau)) \leq \Kc &+ \left({\bf B}^*
(\tau)-{\boldsymbol \mu}^*(\tau)\right)^\transp\left({\bf Q}(jT)+(\tau-jT)\kappav\right) \notag \\ 
&+\left(\gammav^*(\tau)-\Dm^*(\tau)\right)^\transp\left(\Thetav(jT)+(\tau-jT)\kappav\right) \notag\\
&-V\sum_{u \in \Uc}\phi_u(\gamma_u^*(\tau)). \label{oneslot-arbit-wiflix}
\end{align}
}
Then, summing (\ref{oneslot-arbit-wiflix}) over $\tau \in \{jT, \ldots, (j+1)T-1\}$, we obtain the $T$-slot Lyapunov drift over the $j$-th frame:
{\small
\begin{align}
 L&({\bf G}((j+1)T))-L({\bf G}(jT))-V \sum_{\tau=jT}^{jT+T-1}\sum_{u \in \Uc}\phi_u(\gamma_u(\tau)) \notag \\
 & \leq \Kc T+\left(\sum \nolimits_{\tau=jT}^{jT+T-1}\left({\bf B}^*
(\tau)-{\boldsymbol \mu}^*(\tau)\right)\right)^\transp{\bf Q}(jT) + \left(\sum \nolimits_{\tau=jT}^{jT+T-1}\left({\bf B}^*
(\tau)-{\boldsymbol \mu}^*(\tau)\right)\left(\tau-jT\right)\right)^\transp
\kappav \notag\\
&~~~~~~~~+\left(\sum \nolimits_{\tau=jT}^{jT+T-1}\left(\gammav^*(\tau)-\Dm^*(\tau)\right)\right)^\transp\Thetav(jT)+\left(\sum \nolimits_{\tau=jT}^{jT+T-1}\left(\gammav^*(\tau)-\Dm^*(\tau)\right)\left(\tau-jT\right)\right)^\transp
\kappav \notag \\
&~~~~~~~~-V \sum \nolimits_{\tau=jT}^{jT+T-1}\sum_{u \in \Uc}\phi_u(\gamma_u^*(\tau)) \label{dppopt-comparison-wiflix}
\end{align}
}
 Using the inequalities ${\bf B}^*(\tau)-{\boldsymbol \mu}^*(\tau)\leq 2\kappav$,~ 
$\gammav^*(\tau)-\Dm^*(\tau)\leq 2\kappav$ in (\ref{dppopt-comparison-wiflix}), we have
{\small
\begin{align}
 L&({\bf G}((j+1)T))-L({\bf G}(jT))-V \sum_{\tau=jT}^{jT+T-1}\sum_{u \in \Uc}\phi_u(\gamma_u(\tau)) \notag \\
 & \leq \Kc T+\left(\sum \nolimits_{\tau=jT}^{jT+T-1}\left({\bf B}^*
(\tau)-{\boldsymbol \mu}^*(\tau)\right)\right)^\transp{\bf Q}(jT) + 2\left(\sum \nolimits_{\tau=jT}^{jT+T-1}\left(\tau-jT\right)\right)\kappav^\transp\kappav \notag\\
&~~~~~~~~+\left(\sum \nolimits_{\tau=jT}^{jT+T-1}\left(\gammav^*(\tau)-\Dm^*(\tau)\right)\right)^\transp\Thetav(jT)+2\left(\sum \nolimits_{\tau=jT}^{jT+T-1}\left(\tau-jT\right)\right)\kappav^\transp\kappav \notag \\
&~~~~~~~~-V\sum \nolimits_{\tau=jT}^{jT+T-1} \sum_{u \in \Uc}\phi_u(\gamma_u^*(\tau))
\end{align} 
}
Using $\kappav^\transp\kappav \leq \frac{\Kc}{2}$, $\sum_{\tau=jT}^{jT+T-1}(\tau-jT)=\frac{T(T-1)}{2}$, we get
{\small
\begin{align}
 L&({\bf G}((j+1)T))-L({\bf G}(jT))-V\sum_{\tau=jT}^{jT+T-1}\sum_{u \in \Uc}\phi_u(\gamma_u(\tau)) \notag \\
 &\leq \Kc T+\Kc T(T-1) +\left(\sum \nolimits_{\tau=jT}^{jT+T-1}\left({\bf B}^*
(\tau)-{\boldsymbol \mu}^*(\tau)\right)\right)^\transp{\bf Q}(jT) \notag \\
&~~~~~~~~+\left(\sum \nolimits_{\tau=jT}^{jT+T-1}\left(\gammav^*(\tau)-\Dm^*(\tau)\right)\right)^\transp\Thetav(jT)-V\sum \nolimits_{\tau=jT}^{jT+T-1} \sum_{u \in \Uc}\phi_u(\gamma_u^*(\tau))\label{common-allpolicies-wiflix}
\end{align}
}
We  now consider the policy comprising of the control actions $\{m_u^*(\tau)\}_{\tau=jT}^{(j+1)T-1}~\forall~u \in \Uc$, $\{\gammav^*(\tau)\}_{\tau=jT}^{(j+1)T-1}$ and $\{(\mu_{hu}^*(\tau))\}_{\tau=jT}^{(j+1)T-1}$, and satisfying the following constraints:\footnote{It is easy to see that such policy is guaranteed to exist 
provided that we allow, without loss of generality, for a virtual video layer of zero quality and zero rate, and in the assumption that, at any slot $t$, 
each user $u$ has at least one link $(h,u) \in \Ec$ with $h \in \Nc(u) \cap \Hc(f_u)$ with peak rate lower bounded by some strictly positive number $C_{\min}$.
This prevents the case where a user gets zero rate for a whole frame of length $T$. This assumption is not restrictive in practice since a user experiencing unacceptably 
poor link quality to all the helpers for a long time interval would be disconnected from the network and its streaming session halted.}
{\small
\begin{align}
&\frac{1}{T} \sum_{\tau=jT}^{(j+1)T-1}\left[
B^*_{f_u}\left(m_u(\tau),\tau\right)-\mu^*_{u}\left(\tau\right)\right] < -\epsilon~\forall ~u \in \Uc \label{primeineq-1-wiflix}  \\
& \frac{1}{T} \sum_{\tau=jT}^{(j+1)T-1}\left[
\gamma^*_u\left(\tau\right)-D_{f_u}^*\left(m_u(\tau), \tau\right)\right] < -\epsilon~\forall~ u \in \Uc \label{primeineq-2-wiflix}
\end{align}
}
where $\epsilon>0$ is arbitrary. 
We plug in the inequalities (\ref{primeineq-1-wiflix}), (\ref{primeineq-2-wiflix}) in 
(\ref{common-allpolicies-wiflix}) and obtain
{\small
\begin{align}
 L&({\bf G}((j+1)T))-L({\bf G}(jT))-V\sum_{\tau=jT}^{jT+T-1}\sum_{u \in \Uc}\phi_u(\gamma_u(\tau)) \notag \\
 & < \Kc T^2 -\epsilon T \sum_{u \in \Uc}Q_{u}(jT) - \epsilon T \sum_{u \in \Uc}\Theta_{u}(jT)-V\sum \nolimits_{\tau=jT}^{jT+T-1} \sum_{u \in \Uc}\phi_u(\gamma_u^*(\tau))
\end{align}
}
Also, considering the fact that for any vector $\gammav = (\gamma_1, \ldots, \gamma_{|\Uc|})$ we have
\begin{align}
\sum_{u \in \Uc}\phi_u(D_u^{\min}) = \phi_{\min} \leq \sum_{u \in \Uc}\phi_u(\gamma_u) \leq \phi_{\max} = \sum_{u \in \Uc}\phi_u(D_u^{\max}),
\end{align} 
we can write:
\begin{align}
 L({\bf G}((j+1)T))-L({\bf G}(jT)) < \Kc T^2 +VT(\phi_{\max}-\phi_{\min}) - \epsilon T \sum_{u \in \Uc}Q_{u}(jT)-\epsilon T \sum_{u \in \Uc}\Theta_{u}(jT) 
\end{align}
Once again using (\ref{bounding1-wiflix}), (\ref{bounding2-wiflix}), we have:
\begin{align}
 L({\bf G}((j+1)T))-L({\bf G}(jT)) < \Kc T^2 &+VT(\phi_{\max}-\phi_{\min}) - \epsilon \sum_{\tau=jT}^{jT+T-1} \sum_{u \in \Uc}Q_{u}(\tau) \notag \\
 &-\epsilon \sum_{\tau=jT}^{jT+T-1} \sum_{u \in \Uc}\Theta_{u}(\tau) + \epsilon  \kappa |\Uc|T(T-1)
\end{align}
Summing the above over the frames $j \in \{0, \ldots, F-1\}$ yields
\begin{align}
L({\bf G}((FT))-L({\bf G}(0)) < \Kc T^2F &+ VFT(\phi_{\max}-\phi_{\min}) - \epsilon \sum_{\tau=0}^{FT-1} \sum_{u \in \Uc}Q_{u}(\tau) \notag \\
& - \epsilon \sum_{\tau=0}^{FT-1} \sum_{u \in \Uc}\Theta_{u}(\tau) + \epsilon  \kappa |\Uc|FT(T-1)
\end{align}
Rearranging and neglecting appropriate terms, we get
\begin{align}
  \frac{1}{FT} \sum_{\tau=0}^{FT-1} \sum_{u \in \Uc}Q_{u}(\tau) 
+ \frac{1}{FT}\sum_{\tau=0}^{FT-1} \sum_{u \in \Uc}\Theta_{u}(\tau) <  \frac{\Kc T}{\epsilon} &+ \frac{V(\phi_{\max}-\phi_{\min})}{\epsilon}+ \frac{L({\bf G}(0))}{\epsilon FT} \notag \\
&+ \kappa |\Uc|(T-1)
\end{align}
Taking limits as $F \rightarrow \infty$
\begin{align}
 \boxed{\lim_{F \rightarrow \infty} \frac{1}{FT} \sum_{\tau=0}^{FT-1}\left( \sum_{u \in \Uc}Q_{u}(\tau) 
+ \sum_{u \in \Uc}\Theta_{u}(\tau)\right) <  \frac{\Kc T}{\epsilon} + \frac{V(\phi_{\max}-\phi_{\min})}{\epsilon}+ \kappa |\Uc|(T-1)} \label{strongstab-arbit-wiflix}
\end{align}
such that (\ref{strongstab-arbit1-wiflix}) is proved.  

We now consider the policy comprising of the decisions which achieves the optimal solution $\phi_j^{\rm opt}$ to the problem (\ref{constraint-wiflix}) -- (\ref{feasactions-trans-arbit-wiflix}). Using (\ref{qstable-arbit-wiflix}) and (\ref{virt-stable-arbit-wiflix}) in (\ref{common-allpolicies-wiflix}), we have 
\begin{align}
 L({\bf G}((j+1)T))-L({\bf G}(jT))-V\sum_{\tau=jT}^{jT+T-1}\sum_{u \in \Uc}\phi_u(\gamma_u(\tau))\leq \Kc T+\Kc T(T-1)-
V T\phi_j^{\rm opt}
\end{align}
Summing this over $j \in \{0, \ldots, F-1\}$, yields
\begin{align}
L({\bf G}((FT))-L({\bf G}(0))-V \sum_{\tau=0}^{FT-1}\sum_{u \in \Uc}\phi_u(\gamma_u(\tau))\leq \Kc T^2F-V T\sum_{j=0}^{F-1}\phi_j^{\rm opt}.
\end{align}
Dividing both sides by $VFT$ and using the fact that $L({\bf G}((FT)) > 0$ , we get 
\begin{align}
\frac{1}{FT}\sum_{\tau=0}^{FT-1}\sum_{u \in \Uc}\phi_u(\gamma_u(\tau)) \geq \frac{1}{F}\sum_{j=0}^{F-1}\phi_j^{\rm opt} - \frac{\Kc T}{V} - \frac{L({\bf G}(0))}{VTF}.
\end{align}
At this point, using Jensen's inequality, the fact that $\phi_u(\cdot)$ is continuous and non-decreasing for all $u \in \Uc$, 
and the fact that the strong stability of the queues (\ref{strongstab-arbit-wiflix}) implies that 
$\lim_{F \rightarrow \infty} \frac{1}{FT} \sum_{\tau=0}^{FT-1}\Theta_{u}(\tau) < \infty ~\forall~ u \in \Uc$,  
which in turns implies that $\overline{\gamma}_u \leq \overline{D}_u~\forall~u \in \Uc$, we arrive at 
\begin{align}
\boxed{\sum_{u \in \Uc}\phi_u\left(\overline{D}_u\right) \geq
 \lim_{F \rightarrow \infty}\frac{1}{F}\sum_{j=0}^{F-1}\phi_j^{\rm opt} - \frac{\Kc T}{V}.}\label{optutil-arbit-wiflix}
\end{align}
such that (\ref{optutil-arbit1-wiflix}) is proved.  

Thus, the utility under the DPP policy is within $O(1/V)$ of the time average of the $\phi_j^{\rm opt}$ utility values that 
can be achieved only  if knowledge of the future states up to a look-ahead of blocks of $T$ slots.  
If $T$ is increased, then the value of $\phi_j^{\rm opt}$ for every frame $j$ improves since we allow a larger 
look-ahead. However, from (\ref{optutil-arbit-wiflix}), we can see that if $T$ is increased, then $V$ can also be increased in order to maintain the same distance from optimality. This yields a corresponding $O(V)$ increase in the queues backlog. 

For the case where the rate function $B_f(m, t)$, the quality function $D_f(m, t)$ and the topology state $\omega(t)$is stationary and ergodic, the time average in the left hand side of (\ref{strongstab-arbit-wiflix}) and in the right hand side of (\ref{optutil-arbit-wiflix})  become ensemble averages because of ergodicity. Thus, we obtain (\ref{utilperf-iid-wiflix}) and (\ref{strongstab-iid-wiflix}). Furthermore, if the network state is i.i.d., we can take $T = 1$ in the above derivation, obtaining 
the bounds given in Corollary \ref{corcor-wiflix}.

\bibliographystyle{IEEEtran}
\bibliography{dilip-ref}

\end{document}